\documentclass[iop]{emulateapj}
\usepackage[utf8x]{inputenc}
\usepackage{graphicx}
\usepackage{amssymb}
\usepackage{amsmath}
\usepackage{amssymb}
\usepackage{amsmath}
\usepackage{color}
\usepackage{times}
\usepackage{url}
\tabletypesize{\scriptsize}

\newcommand{\msol}{\textrm{M}_{\odot}}

\newcommand{\civ}{C{\footnotesize IV}}
\newcommand{\mgii}{Mg{\footnotesize II}}

\usepackage{eso-pic}% http://ctan.org/pkg/eso-pic

\AddToShipoutPictureBG*{%
  \AtPageUpperLeft{%
    \hspace{0.75\paperwidth}%
    \raisebox{-3.5\baselineskip}{%
      \makebox[0pt][l]{\textnormal{DES 2017-0257}}
}}}%

\AddToShipoutPictureBG*{%
  \AtPageUpperLeft{%
    \hspace{0.75\paperwidth}%
    \raisebox{-4.5\baselineskip}{%
      \makebox[0pt][l]{\textnormal{Fermilab PUB-17-506-E}}
}}}%

\makeatletter
\def\@to{to}
\makeatother

\begin{document}

\title{Quasar Accretion Disk Sizes From Continuum Reverberation Mapping From the Dark Energy Survey}

\author{D.~Mudd}
\affiliation{Department of Physics and Astronomy, University of California Irvine, Irvine, CA 92697}
\affiliation{Department of Astronomy, The Ohio State University, Columbus, OH 43210, USA}
\email{dmmudd@uci.edu}
\author{P.~Martini}
\affiliation{Center for Cosmology and Astro-Particle Physics, The Ohio State University, Columbus, OH 43210, USA}
\affiliation{Department of Astronomy, The Ohio State University, Columbus, OH 43210, USA}
\author{Y.~Zu}
\affiliation{Center for Cosmology and Astro-Particle Physics, The Ohio State University, Columbus, OH 43210, USA}
\affiliation{Department of Astronomy, The Ohio State University, Columbus, OH 43210, USA}
\author{C.~Kochanek}
\affiliation{Department of Astronomy, The Ohio State University, Columbus, OH 43210, USA}
\author{B.~M.~Peterson}
\affiliation{Center for Cosmology and Astro-Particle Physics, The Ohio State University, Columbus, OH 43210, USA}
\affiliation{Department of Astronomy, The Ohio State University, Columbus, OH 43210, USA}
\affiliation{Space Telescope Science Institute, 3700 San Martin Drive, Baltimore, MD 21218}
\author{R.~Kessler}
\affiliation{Kavli Institute for Cosmological Physics, University of Chicago, Chicago, IL 60637, USA}
\author{T.~M.~Davis}
\affiliation{School of Mathematics and Physics, University of Queensland,  Brisbane, QLD 4072, Australia}
\affiliation{ARC Centre of Excellence for All-sky Astrophysics (CAASTRO)}
\author{J.~K.~Hoormann}
\affiliation{School of Mathematics and Physics, University of Queensland,  Brisbane, QLD 4072, Australia}
\author{A.~King}
\affiliation{School of Mathematics and Physics, University of Queensland,  Brisbane, QLD 4072, Australia}
\author{C.~Lidman}
\affiliation{Australian Astronomical Observatory, North Ryde, NSW 2113, Australia}
\affiliation{ARC Centre of Excellence for All-sky Astrophysics (CAASTRO)}
\author{N.~E.~Sommer}
\affiliation{The Research School of Astronomy and Astrophysics, Australian National University, ACT 2601, Australia}
\affiliation{ARC Centre of Excellence for All-sky Astrophysics (CAASTRO)}
\author{B.~E.~Tucker}
\affiliation{The Research School of Astronomy and Astrophysics, Australian National University, ACT 2601, Australia}
\affiliation{ARC Centre of Excellence for All-sky Astrophysics (CAASTRO)}
\author{J.~Asorey}
\affiliation{ARC Centre of Excellence for All-sky Astrophysics (CAASTRO)}
\affiliation{Centre for Astrophysics \& Supercomputing, Swinburne University of Technology, Victoria 3122, Australia}
\affiliation{School of Mathematics and Physics, University of Queensland,  Brisbane, QLD 4072, Australia}
\author{S.~Hinton}
\affiliation{School of Mathematics and Physics, University of Queensland,  Brisbane, QLD 4072, Australia}
\affiliation{ARC Centre of Excellence for All-sky Astrophysics (CAASTRO)}
\author{K.~Glazebrook}
\affiliation{Centre for Astrophysics and Supercomputing, Swinburne University of Technology, Hawthorn, VIC 3122, Australia}
\author{K.~Kuehn}
\affiliation{Australian Astronomical Observatory, North Ryde, NSW 2113, Australia}
\author{G.~Lewis}
\affiliation{Sydney Institute for Astronomy, School of Physics, A28, The University of Sydney, NSW 2006, Australia}
\author{E.~Macaulay}
\affiliation{Institute of Cosmology \& Gravitation, University of Portsmouth, Portsmouth PO1 3FX, UK}
\author{A.~Moeller}
\affiliation{The Research School of Astronomy and Astrophysics, Australian National University, ACT 2601, Australia}
\affiliation{ARC Centre of Excellence for All-sky Astrophysics (CAASTRO)}
\author{C.~O'Neill}
\affiliation{School of Mathematics and Physics, University of Queensland,  Brisbane, QLD 4072, Australia}
\affiliation{ARC Centre of Excellence for All-sky Astrophysics (CAASTRO)}
\author{B.~Zhang}
\affiliation{The Research School of Astronomy and Astrophysics, Australian National University, ACT 2601, Australia}
\affiliation{ARC Centre of Excellence for All-sky Astrophysics (CAASTRO)}
\author{T.~M.~C.~Abbott}
\affiliation{Cerro Tololo Inter-American Observatory, National Optical Astronomy Observatory, Casilla 603, La Serena, Chile}
\author{F.~B.~Abdalla}
\affiliation{Department of Physics and Electronics, Rhodes University, PO Box 94, Grahamstown, 6140, South Africa}
\affiliation{Department of Physics \& Astronomy, University College London, Gower Street, London, WC1E 6BT, UK}
\author{S.~Allam}
\affiliation{Fermi National Accelerator Laboratory, P. O. Box 500, Batavia, IL 60510, USA}
\author{M.~Banerji}
\affiliation{Kavli Institute for Cosmology, University of Cambridge, Madingley Road, Cambridge CB3 0HA, UK}
\affiliation{Institute of Astronomy, University of Cambridge, Madingley Road, Cambridge CB3 0HA, UK}
\author{A.~Benoit-L{\'e}vy}
\affiliation{Sorbonne Universit\'es, UPMC Univ Paris 06, UMR 7095, Institut d'Astrophysique de Paris, F-75014, Paris, France}
\affiliation{CNRS, UMR 7095, Institut d'Astrophysique de Paris, F-75014, Paris, France}
\affiliation{Department of Physics \& Astronomy, University College London, Gower Street, London, WC1E 6BT, UK}
\author{E.~Bertin}
\affiliation{CNRS, UMR 7095, Institut d'Astrophysique de Paris, F-75014, Paris, France}
\affiliation{Sorbonne Universit\'es, UPMC Univ Paris 06, UMR 7095, Institut d'Astrophysique de Paris, F-75014, Paris, France}
\author{D.~Brooks}
\affiliation{Department of Physics \& Astronomy, University College London, Gower Street, London, WC1E 6BT, UK}
\author{A.~Carnero~Rosell}
\affiliation{Laborat\'orio Interinstitucional de e-Astronomia - LIneA, Rua Gal. Jos\'e Cristino 77, Rio de Janeiro, RJ - 20921-400, Brazil}
\affiliation{Observat\'orio Nacional, Rua Gal. Jos\'e Cristino 77, Rio de Janeiro, RJ - 20921-400, Brazil}
\author{D.~Carollo}
\affiliation{INAF - Osservatorio Astrofisico di Torino, Pino Torinese, Italy}
\affiliation{ARC Centre of Excellence for All-sky Astrophysics (CAASTRO)}
\author{M.~Carrasco~Kind}
\affiliation{Department of Astronomy, University of Illinois, 1002 W. Green Street, Urbana, IL 61801, USA}
\affiliation{National Center for Supercomputing Applications, 1205 West Clark St., Urbana, IL 61801, USA}
\author{J.~Carretero}
\affiliation{Institut de F\'{\i}sica d'Altes Energies (IFAE), The Barcelona Institute of Science and Technology, Campus UAB, 08193 Bellaterra (Barcelona) Spain}
\author{C.~E.~Cunha}
\affiliation{Kavli Institute for Particle Astrophysics \& Cosmology, P. O. Box 2450, Stanford University, Stanford, CA 94305, USA}
\author{C.~B.~D'Andrea}
\affiliation{Department of Physics and Astronomy, University of Pennsylvania, Philadelphia, PA 19104, USA}
\author{L.~N.~da Costa}
\affiliation{Laborat\'orio Interinstitucional de e-Astronomia - LIneA, Rua Gal. Jos\'e Cristino 77, Rio de Janeiro, RJ - 20921-400, Brazil}
\affiliation{Observat\'orio Nacional, Rua Gal. Jos\'e Cristino 77, Rio de Janeiro, RJ - 20921-400, Brazil}
\author{C.~Davis}
\affiliation{Kavli Institute for Particle Astrophysics \& Cosmology, P. O. Box 2450, Stanford University, Stanford, CA 94305, USA}
\author{S.~Desai}
\affiliation{Department of Physics, IIT Hyderabad, Kandi, Telangana 502285, India}
\author{P.~Doel}
\affiliation{Department of Physics \& Astronomy, University College London, Gower Street, London, WC1E 6BT, UK}
\author{P.~Fosalba}
\affiliation{Institute of Space Sciences, IEEC-CSIC, Campus UAB, Carrer de Can Magrans, s/n,  08193 Barcelona, Spain}
\author{J.~Garc\'ia-Bellido}
\affiliation{Instituto de Fisica Teorica UAM/CSIC, Universidad Autonoma de Madrid, 28049 Madrid, Spain}
\author{E.~Gaztanaga}
\affiliation{Institute of Space Sciences, IEEC-CSIC, Campus UAB, Carrer de Can Magrans, s/n,  08193 Barcelona, Spain}
\author{D.~W.~Gerdes}
\affiliation{Department of Physics, University of Michigan, Ann Arbor, MI 48109, USA}
\affiliation{Department of Astronomy, University of Michigan, Ann Arbor, MI 48109, USA}
\author{D.~Gruen}
\affiliation{SLAC National Accelerator Laboratory, Menlo Park, CA 94025, USA}
\affiliation{Kavli Institute for Particle Astrophysics \& Cosmology, P. O. Box 2450, Stanford University, Stanford, CA 94305, USA}
\author{R.~A.~Gruendl}
\affiliation{National Center for Supercomputing Applications, 1205 West Clark St., Urbana, IL 61801, USA}
\affiliation{Department of Astronomy, University of Illinois, 1002 W. Green Street, Urbana, IL 61801, USA}
\author{J.~Gschwend}
\affiliation{Observat\'orio Nacional, Rua Gal. Jos\'e Cristino 77, Rio de Janeiro, RJ - 20921-400, Brazil}
\affiliation{Laborat\'orio Interinstitucional de e-Astronomia - LIneA, Rua Gal. Jos\'e Cristino 77, Rio de Janeiro, RJ - 20921-400, Brazil}
\author{G.~Gutierrez}
\affiliation{Fermi National Accelerator Laboratory, P. O. Box 500, Batavia, IL 60510, USA}
\author{W.~G.~Hartley}
\affiliation{Department of Physics \& Astronomy, University College London, Gower Street, London, WC1E 6BT, UK}
\affiliation{Department of Physics, ETH Zurich, Wolfgang-Pauli-Strasse 16, CH-8093 Zurich, Switzerland}
\author{K.~Honscheid}
\affiliation{Center for Cosmology and Astro-Particle Physics, The Ohio State University, Columbus, OH 43210, USA}
\affiliation{Department of Physics, The Ohio State University, Columbus, OH 43210, USA}
\author{D.~J.~James}
\affiliation{Astronomy Department, University of Washington, Box 351580, Seattle, WA 98195, USA}
\author{S.~Kuhlmann}
\affiliation{Argonne National Laboratory, 9700 South Cass Avenue, Lemont, IL 60439, USA}
\author{N.~Kuropatkin}
\affiliation{Fermi National Accelerator Laboratory, P. O. Box 500, Batavia, IL 60510, USA}
\author{M.~Lima}
\affiliation{Laborat\'orio Interinstitucional de e-Astronomia - LIneA, Rua Gal. Jos\'e Cristino 77, Rio de Janeiro, RJ - 20921-400, Brazil}
\affiliation{Departamento de F\'isica Matem\'atica, Instituto de F\'isica, Universidade de S\~ao Paulo, CP 66318, S\~ao Paulo, SP, 05314-970, Brazil}
\author{M.~A.~G.~Maia}
\affiliation{Laborat\'orio Interinstitucional de e-Astronomia - LIneA, Rua Gal. Jos\'e Cristino 77, Rio de Janeiro, RJ - 20921-400, Brazil}
\affiliation{Observat\'orio Nacional, Rua Gal. Jos\'e Cristino 77, Rio de Janeiro, RJ - 20921-400, Brazil}
\author{J.~L.~Marshall}
\affiliation{George P. and Cynthia Woods Mitchell Institute for Fundamental Physics and Astronomy, and Department of Physics and Astronomy, Texas A\&M University, College Station, TX 77843,  USA}
\author{R.~G.~McMahon}
\affiliation{Institute of Astronomy, University of Cambridge, Madingley Road, Cambridge CB3 0HA, UK}
\affiliation{Kavli Institute for Cosmology, University of Cambridge, Madingley Road, Cambridge CB3 0HA, UK}
\author{F.~Menanteau}
\affiliation{Department of Astronomy, University of Illinois, 1002 W. Green Street, Urbana, IL 61801, USA}
\affiliation{National Center for Supercomputing Applications, 1205 West Clark St., Urbana, IL 61801, USA}
\author{R.~Miquel}
\affiliation{Institut de F\'{\i}sica d'Altes Energies (IFAE), The Barcelona Institute of Science and Technology, Campus UAB, 08193 Bellaterra (Barcelona) Spain}
\affiliation{Instituci\'o Catalana de Recerca i Estudis Avan\c{c}ats, E-08010 Barcelona, Spain}
\author{A.~A.~Plazas}
\affiliation{Jet Propulsion Laboratory, California Institute of Technology, 4800 Oak Grove Dr., Pasadena, CA 91109, USA}
\author{A.~K.~Romer}
\affiliation{Department of Physics and Astronomy, Pevensey Building, University of Sussex, Brighton, BN1 9QH, UK}
\author{E.~Sanchez}
\affiliation{Centro de Investigaciones Energ\'eticas, Medioambientales y Tecnol\'ogicas (CIEMAT), Madrid, Spain}
\author{R.~Schindler}
\affiliation{SLAC National Accelerator Laboratory, Menlo Park, CA 94025, USA}
\author{M.~Schubnell}
\affiliation{Department of Physics, University of Michigan, Ann Arbor, MI 48109, USA}
\author{M.~Smith}
\affiliation{School of Physics and Astronomy, University of Southampton,  Southampton, SO17 1BJ, UK}
\author{R.~C.~Smith}
\affiliation{Cerro Tololo Inter-American Observatory, National Optical Astronomy Observatory, Casilla 603, La Serena, Chile}
\author{M.~Soares-Santos}
\affiliation{Fermi National Accelerator Laboratory, P. O. Box 500, Batavia, IL 60510, USA}
\author{F.~Sobreira}
\affiliation{Instituto de F\'isica Gleb Wataghin, Universidade Estadual de Campinas, 13083-859, Campinas, SP, Brazil}
\affiliation{Laborat\'orio Interinstitucional de e-Astronomia - LIneA, Rua Gal. Jos\'e Cristino 77, Rio de Janeiro, RJ - 20921-400, Brazil}
\author{E.~Suchyta}
\affiliation{Computer Science and Mathematics Division, Oak Ridge National Laboratory, Oak Ridge, TN 37831}
\author{M.~E.~C.~Swanson}
\affiliation{National Center for Supercomputing Applications, 1205 West Clark St., Urbana, IL 61801, USA}
\author{G.~Tarle}
\affiliation{Department of Physics, University of Michigan, Ann Arbor, MI 48109, USA}
\author{D.~Thomas}
\affiliation{Institute of Cosmology \& Gravitation, University of Portsmouth, Portsmouth, PO1 3FX, UK}
\author{D.~L.~Tucker}
\affiliation{Fermi National Accelerator Laboratory, P. O. Box 500, Batavia, IL 60510, USA}
\author{A.~R.~Walker}
\affiliation{Cerro Tololo Inter-American Observatory, National Optical Astronomy Observatory, Casilla 603, La Serena, Chile}

\collaboration{DES Collaboration}

\shorttitle{DES Quasar Accretion Disk Sizes}

\begin{abstract}
We present accretion disk size measurements for 15 luminous quasars at $0.7 \leq z \leq 1.9$ derived from $griz$ light curves from the Dark Energy Survey. We measure the disk sizes with continuum reverberation mapping using two methods, both of which are derived from the expectation that accretion disks have a radial temperature gradient and the continuum emission at a given radius is well-described by a single blackbody. In the first method we measure the relative lags between the multiband light curves, which provides the relative time lag between shorter and longer wavelength variations. From this, we are only able to constrain upper limits on disk sizes, as many are consistent with no lag the 2$\sigma$ level.  The second method fits the model parameters for the canonical thin disk directly rather than solving for the individual time lags between the light curves. Our measurements demonstrate good agreement with the sizes predicted by this model for accretion rates between 0.3-1 times the Eddington rate. Given our large uncertainties, our measurements are also consistent with disk size measurements from gravitational microlensing studies of strongly lensed quasars, as well as other photometric reverberation mapping results, that find disk sizes that are a factor of a few ($\sim$3) larger than predictions.
\end{abstract}

\keywords{galaxies: active, accretion disks, quasars: general}

\section{Introduction}
\label{sec: intro}
The spectra of quasars and less-luminous active galactic nuclei (AGN) are characterized by the presence of both narrow emission lines that originate far from the central engine and broad ones that originate close to the supermassive black hole (SMBH).  These lines are superposed on a continuum mostly due to thermal radiation from the accretion disk that peaks in the ultraviolet.  Understanding the size and structure of this accretion disk is important because it is energetically dominant, drives most of the other emission, and probes the growth of the central SMBH. 

The canonical quasar accretion disk model is the optically thick, geometrically thin disk \citep{Lynden-Bell69, Shakura73}.  The disk emission is a combination of the local thermal emission of the viscously dissipated energy and reprocessing of emission from the inner regions.  Thin disks are stable at low to moderate accretion rates compared to the Eddington rate.  There are also stable slim accretion disks \citep{Abramowicz88, Narayan95}, where the optically thin disks are no longer geometrically thin and the accretion rates are near or super-Eddington.  \citet{Hall17} propose an extension to the thin disk model for the accretion where the emission is modified by a low density disk atmosphere to be non-thermal, making disks appear to be larger than would be inferred from a black body.

Viscously-heated material moving through the accretion disk at radii larger than the innermost stable circular orbit emits photons as
\begin{eqnarray}
\sigma T^{4} &=& \frac{3GM\dot{m}}{8\pi R^{3}},
\end{eqnarray}
where $T$ is the temperature of the material, $M$ is the mass of black hole, $\dot{m}$ is the accretion rate onto the black hole, and $R$ is the orbital radius of the material \citep{Collier98}.  The size, R, of this standard thin disk at an effective emitting wavelength $\lambda_{0}$, defined by the region where the disk temperature is $kT = hc/\lambda_{0}$, and assuming the annulus is emitting as a blackbody \citep{Collier98, Morgan10}, can be rewritten as
\begin{eqnarray}
\label{eqnarray: thin_disk_Kochanek}
 R_{\lambda_{0}} &=& \left[\frac{45G\lambda_{0}^{4}M\dot{m}(3 + \kappa)}{16\pi^{6}hc^{2}}\right]^{1/3} \\
                 &=& 9.7 \times 10^{15}\left(\frac{\lambda_{0}}{\mu\textrm{m}}\right)^{\beta}\left(\frac{M}{10^{9}\msol}\right)^{2/3}\left(\frac{L}{\eta L_{\textrm{E}}}\right)^{1/3} \textrm{ cm}, \nonumber
\end{eqnarray}
where $M$ is again the black hole mass, $\dot{m}$ is the accretion rate, $\kappa$ is the ratio of external to local radiative heating, $L_{\textrm{E}}$ is the Eddington luminosity, and $\beta$ is the power law index on $\lambda$, which would be 4/3 from Equation \ref{eqnarray: thin_disk_Kochanek}.  We introduce this $\beta$ here as we will use it throughout the paper.  The above scaling relationship assumes that $\kappa$ = 0.  The accretion rate can be related to the luminosity by $L = \eta\dot{m}c^{2}$, where $\eta$ is the radiative efficiency of converting the accreted mass into energy.  From Equation \ref{eqnarray: thin_disk_Kochanek}, we find that the accretion disk size should scale as $\lambda^{4/3}$.  Given the blackbody emission assumption that $kT = hc/\lambda_{0}$, this predicts that the temperature also goes as $T \propto R_{\lambda_{0}}^{-1/\beta} \propto R_{\lambda_{0}}^{-3/4}$.  We generalize the disk radius at an emitting wavelength $\lambda$ as

\begin{equation}
 R_{\lambda} = R_{\lambda_{0}}\left(\frac{\lambda}{\lambda_{0}}\right)^{\beta}
 \label{eq: thin_disk_short}
\end{equation}
where the standard thin disk has $R_{\lambda_{0}}$ as given in Equation \ref{eqnarray: thin_disk_Kochanek} and $\beta = 4/3$ is the prediction for infalling viscously-heated material.

If the temporal variability of the disk is driven by fluctuations in the irradiated flux from a central source, such as in the ``lamppost'' model (see \citealp{Cackett07} and references therein), then the time delay for light to reach any radius in the disk implies that there should be delays in the response at different wavelengths.  If the size-wavelength relation is given by Equation \ref{eq: thin_disk_short}, then the lag $\tau$ between emission from different annuli in the disk with effective emitting wavelengths $\lambda_{0}$ and $\lambda$ should be of the form
\begin{equation}
 \tau = \frac{R_{\lambda}}{c}\left[\left(\frac{\lambda}{\lambda_{0}}\right)^{\beta} - 1\right].
 \label{eq: thin_disk_lag}
\end{equation}

The above derivations all assume (by making $kT = hc/\lambda_{0}$) that the observed emission is tied uniquely to a single emitting annulus.  In reality, each annulus emits as a blackbody, not at a single wavelength, so one observed wavelength corresponds to the combined emission from a suite of annuli.  To get the effective flux-weighted mean emitting radius for any given wavelength, integrating over the surface brightness profile of the disk $B(T(R))$ at $R$ from some inner edge $R_{0}$ gives
\begin{eqnarray}
<R_{\lambda_{0}}> = \frac{\int_{R_{0}}^{\infty} B(T(R))R^{2} dR}{\int_{R_{0}}^{\infty} B(T(R))R dR}.
\label{eq: r_mean}
\end{eqnarray}
We account for this overlapping of fluxes by stating that the effective temperature at $\lambda_0$ is given by $kT=Xhc/\lambda_0$, where $X$ takes into account the range of different annuli sizes that contribute to the emission at wavelength $\lambda_0$.  Propagating this change through the analysis above and allowing for $\kappa$ to be nonzero, we have
\begin{eqnarray}
R_{\lambda} = \frac{1}{c}\left(\frac{Xhc}{k_{B}T}\right)^{4/3}\left[\left(\frac{GM}{8\pi\sigma}\right)\left(\frac{L_{Edd}}{\eta c^{2}}\right)\left(3 + \kappa\right)\dot{m}\right]^{1/3},
\label{eqn: thin_disk_Fausnaugh}
\end{eqnarray}
where now $R_{\lambda}$ is this flux-weighted emitting radius.  By performing the integral in Equation \ref{eq: r_mean} and dividing this radius by the prediction from Equation \ref{eqnarray: thin_disk_Kochanek}, one finds 
\begin{eqnarray}
X = \frac{<R_{\lambda_{0}}>}{R_{\lambda_{0}}}.
\label{eqn: X}
\end{eqnarray}
This corrective X factor depends on both the wavelength of emission and the black hole mass. It is typically of order unity, and was 2.49 for NGC 5548 for $\lambda_{0} = 1367\AA$ \citep{Fausnaugh16}.  

A number of quasar accretion disk sizes have been measured using gravitational microlensing.  In quasar microlensing, the amplitude of the variability encodes the disk size (see, e.g., the review by \citealp{Wambsganss01}).  At optical wavelengths, disk sizes appear to follow the $M_{\textrm{BH}}^{2/3}$ scaling expected for thin disks, but are larger in absolute size than predicted from Equation \ref{eqnarray: thin_disk_Kochanek} by a factor of 2-3 in some cases \citep{Morgan10}.  There is also evidence from lensed quasars that the temperature profile $\beta$ may be smaller than the thin disk prediction.  \citet{Jimenez-Vicente14} found that the best-fit profile was 0.8.  While it will be feasible to expand these studies to hundreds of lenses in the era of the Large Synoptic Survey Telescope, lensed systems will always be a relatively rare subset of quasars.  

An alternative technique for measuring accretion disk sizes is reverberation mapping \citep{Blandford82, Peterson93, Wanders97, Collier98, Kriss00, Sergeev05}.  This was developed to measure the distance to the broad line region by observing the time delay between variability in the continuum emission (as a proxy for the ionizing radiation from the innermost regions of the disk) and the response of the broad emission lines.  This method can also be adapted to measure the relative time lag between the continuum emission at two wavelengths, e.g. two photometric bandpasses, which is then a proxy for the difference in the disk radii contributing to the emission (Equation \ref{eq: thin_disk_short}).  This allows for estimating effective accretion disk sizes at those emitting wavelengths in unlensed quasars, and thus can be done on a much larger sample of objects.  The primary challenge is that quasars show little variability on the short timescales corresponding to light travel times across the disk (e.g., \citealp{MacLeod10}).

Early attempts at disk reverberation mapping include \citet{Wanders97} and \citet{Collier98} for NGC 7469 and \citet{Peterson98} for NGC 4151.  More recently, \citet{Sergeev05} measured continuum lags at 2$\sigma$ or upper limits for approximately a dozen objects.  Interpreting the lags as light travel delays across the disk, the lags implied disk sizes growing as $L^{0.4-0.5}$, close to their prediction of $L^{0.5}$ from lags due to simple radiative propagation of variability \citep{Sergeev05}.  \citet{Shappee15} and \citet{Fausnaugh16} observed a wide range of wavelengths tracking the broadband variability of NGC 2617 and NGC 5548 (respectively) from the X-rays through the infrared.  Both of these studies found the wavelength dependence of their accretion disk fits to be consistent with the thin-disk prediction of $\beta =$4/3, with evidence that it may be closer to $\beta$=1.  Like some of the microlensing results, the accretion disk size estimates from these two studies were larger than predicted given the black hole mass estimates for these objects.  In particular, the Space Telescope and Optical Reverberation Mapping (AGN STORM) campaign for NGC 5548 (e.g., \citealp{DeRosa15}) is the highest quality variability dataset for a single object, and \citet{Edelson15, Fausnaugh16} conclude that the accretion disk is roughly three times larger than expected from Equation \ref{eqn: thin_disk_Fausnaugh}, assuming an accretion at 10\% the Eddington rate.  The Pan-STARRS collaboration performed a similar analysis for the \emph{griz} bands on a sample of higher luminosity quasars.  They restricted their work to two redshift bins that minimized broad emission line contamination to the optical filters, and consistently find that the accretion disks are larger than expected from thin disk theory \citep{Jiang17}.

In this paper, we use data from the Dark Energy Survey (DES; \citealp{Flaugher05, DES16}) to detect continuum time delays and study accretion disks in a sample of $z \geq 0.7$ quasars.  In Section \ref{sec: observations}, we briefly describe the DES survey and data.  In Section \ref{sec: analysis}, we describe the analysis of our data and our methodologies.  We conclude in Section \ref{sec: conc}.  

\section{Observations}
\label{sec: observations}
DES is a five-year, five band, optical, photometric survey covering 5000 square degrees of the southern sky, starting in 2013, using the Dark Energy Camera \citep{Flaugher15} on the Victor M. Blanco 4m telescope near Cerro Tololo Inter-American Observatory near La Serena, Chile.  The primary goal of the survey is to investigate the expansion of the universe using weak lensing, baryon acoustic oscillations, galaxy clusters, and Type Ia supernovae \citep{Flaugher05, Flaugher15}.  As part of the supernova search, 30 square degrees are be repeatedly observed on an approximately weekly cadence in the \emph{g, r, i,} and \emph{z} filters for the duration of DES, amounting to 20-30 epochs per filter per year \citep{Kessler15}.  In these fields, we also have a set of spectroscopically verified quasars from the OzDES program (Australian DES/Optical redshifts for DES; \citealp{Yuan15, Childress17}) as part of the DES and OzDES reverberation mapping project.  This project photometrically and spectroscopically monitors 771 quasars in order to measure the masses of their supermassive black holes \citep{King15}.  These quasars are as faint as 23.6 in \emph{g}, and were selected heterogeneously with a broad range of quasar detection techniques (e.g., \citealp{Banerji15}, \citealp{Tie17}).  From the combination of these two surveys, we use the first three years of photometry and spectra for the 771 quasars that we continue to monitor \citep{Diehl16, Childress17}.

\section{Analysis}
\label{sec: analysis}
We used the image subtraction pipeline developed for the analysis of supernova light curves \citep{Kessler15, Goldstein15} to create our quasar light curves.  We visually inspected the light curves for each of the approximately 800 reverberation mapping quasars in the DES SN fields that are part of the spectroscopic reverberation mapping project to look for good continuum variability candidates.  We constructed our final list of 15 candidates as the subset of all quasars that had a substantial flux variation ($\geq$5 times the photometric errors) on a timescale of weeks in both the \emph{g} and \emph{z} bands, excluding the first or last few data points of each observing season.  This was to avoid variations that would occur in other bands that would require data before or after the DES observing season.  These bands provide the longest wavelength baseline and represent the maximum (\emph{g}) and minimum (\emph{z}) levels of expected variability.  Significant variability in both bands gives the best indication that we can recover a time delay.

Table \ref{tab: sample} summarizes the general properties of the quasars in this final sample of 15 quasars that were classified as our best photometric lag candidates for having the highest photometric variability and multi-band cadence around the feature(s).  We derive lag results from the observations in the season in which the highest amplitude variability was observed.  Five of the quasars exhibited strong variability in more than one season, and three showed it in all three seasons.  For these sources, we analyze each season independently.  The light curves, shown in Figure \ref{fig: LCExample} and as Table \ref{tab: phot_table}, have on average 30 epochs per season per band and an average cadence of 6-8 nights between observations.    All four bands are typically observed on the same night.  We also note that one of our quasars, DES J024854.79+001054.12, is separated from a FIRST radio source by $0.^{\!\!\prime\prime}9$, and therefore some of its variability may arise from a jet rather than changes in the accretion disk.  In some cases we do not analyze multiple seasons because individual seasons show no evidence of variability.

\begin{deluxetable}{lrrrrr}
\tablecaption{ Sample Description } 
\tablewidth{0pt}
\tabletypesize{\tiny}
\tablewidth{0.40\textwidth}
\tablehead{
\colhead{Quasar Name} & 
\colhead{RA} & 
\colhead{Dec} & 
\colhead{z} & 
\colhead{SN} & 
\colhead{BH Mass} \\
\colhead{} & 
\colhead{} & 
\colhead{} & 
\colhead{} & 
\colhead{Field} & 
\colhead{$10^{9}\textrm{M}_{\odot}$}
}
\startdata
DES J025318.76+000414.20 &  43.32817 &   0.07061 &   1.56 &   S1 & 0.49  \\
DES J024753.20-002137.69 &  41.97167 &  -0.36047 &   1.44 &   S1 & 1.17  \\
DES J021500.22-043007.49 &  33.75092 &  -4.50208 &   1.01 &   X1 & 0.38  \\
DES J022440.70-043657.60 &  36.16960 &  -4.61600 &   0.91 &   X3 & 0.20  \\
DES J022436.17-065912.30 &  36.15071 &  -6.98675 &   1.36 &   X2 & 0.52  \\
DES J022108.60-061753.20 &  35.28583 &  -6.29811 &   1.22 &   X2 & 0.77  \\
DES J022344.56-064039.00 &  35.93568 &  -6.67750 &   0.98 &   X2 & 0.06  \\
DES J033719.99-262418.83 &  54.33328 & -26.40523 &   1.17 &   C1 & 1.01  \\
DES J024918.24-001730.98 &  42.32600 &  -0.29194 &   1.24 &   S1 & 0.49  \\
DES J024854.79+001054.12 &  42.22830 &   0.18170 &   1.15 &   S1 & 0.90  \\
DES J024133.65-010724.20 &  40.39021 &  -1.12339 &   1.87 &   S2 & 0.79  \\
DES J024159.74-010512.41 &  40.49892 &  -1.08678 &   0.90 &   S2 & 0.61  \\
DES J021514.27-053321.31 &  33.80946 &  -5.55592 &   0.70 &   X1 & 0.04  \\
DES J024357.90-011330.40 &  40.99125 &  -1.22511 &   0.90 &   S2 & 0.14  \\
DES J021952.14-040919.91 &  34.96725 &  -4.15553 &   0.69 &   X1 & 0.66  \\
\enddata
\label{tab: sample}
\tablecomments{DES quasar sample analyzed in this work.  ``SN Field'' denotes the supernova field in which the quasar resides \citep{Flaugher05, Flaugher15}.  Note that all masses have an uncertainty of 0.4 dex, calculated using the \citet{McLure02} relationship for MgII.}
\end{deluxetable}

\begin{deluxetable}{lrrrr}
\tablecaption{ DES Quasar Light Curves } 
\tablewidth{0pt}
\tabletypesize{\tiny}
\tablewidth{0.48\textwidth}
\tablehead{
\colhead{Quasar Name} & 
\colhead{MJD} & 
\colhead{Mag} & 
\colhead{Mag$_{\textrm{Err}}$} & 
\colhead{Band} \\ 
}
\startdata
 DES J024133.65-010724.20 & 56534.283 &  20.89 &  0.02 & g \\
 DES J024133.65-010724.20 & 56538.325 &  20.91 &  0.02 & g \\
 DES J024133.65-010724.20 & 56543.296 &  20.86 &  0.02 & g \\
 DES J024133.65-010724.20 & 56547.226 &  20.93 &  0.02 & g \\
 DES J024133.65-010724.20 & 56550.246 &  20.78 &  0.06 & g \\
 DES J024133.65-010724.20 & 56559.205 &  20.96 &  0.05 & g \\
 DES J024133.65-010724.20 & 56567.171 &  21.13 &  0.03 & g \\
 DES J024133.65-010724.20 & 56579.139 &  21.06 &  0.04 & g \\
\enddata
\label{tab: phot_table}
\tablecomments{Photometry for the DES quasars on which we perform our analyses.  All magnitudes are in the AB System.  The full table is available on the online version.}
\end{deluxetable}

\begin{figure*}
  \centerline{%
    \includegraphics[width = 8in]{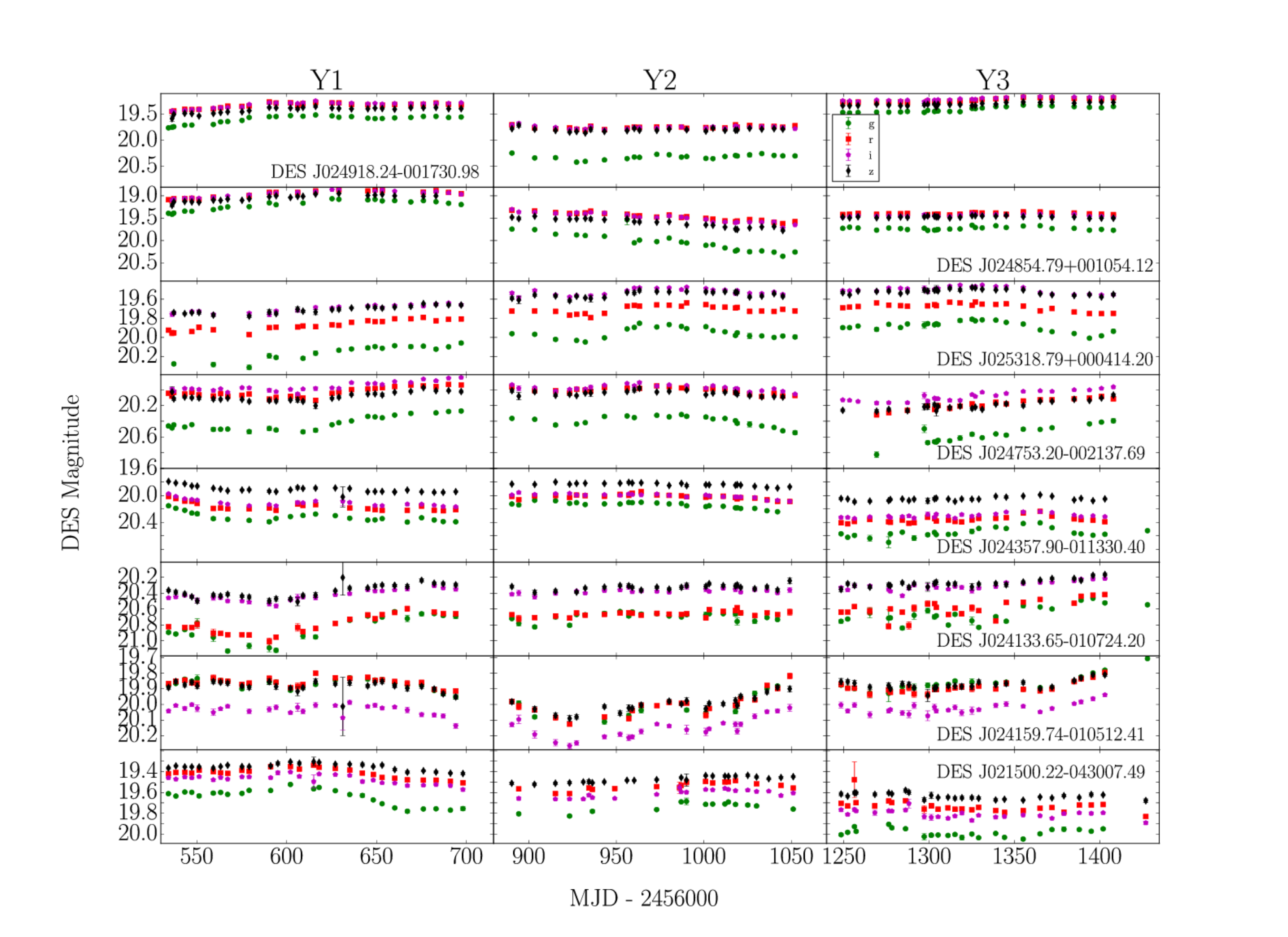}
}%
\caption{DES light curves for the 15 quasars in our sample.  All of these objects show strong variability in at least one observing season.  The DES Y1, Y2, and Y3 data are in the left, middle, and right columns, respectively.  The photometry uses the DES supernova search image subtraction pipeline \citep{Kessler15}.}
  \label{fig: LCExample}
\end{figure*}
\begin{figure*}%ContinuedFloat%
  \centerline{%
    \includegraphics[width = 8in]{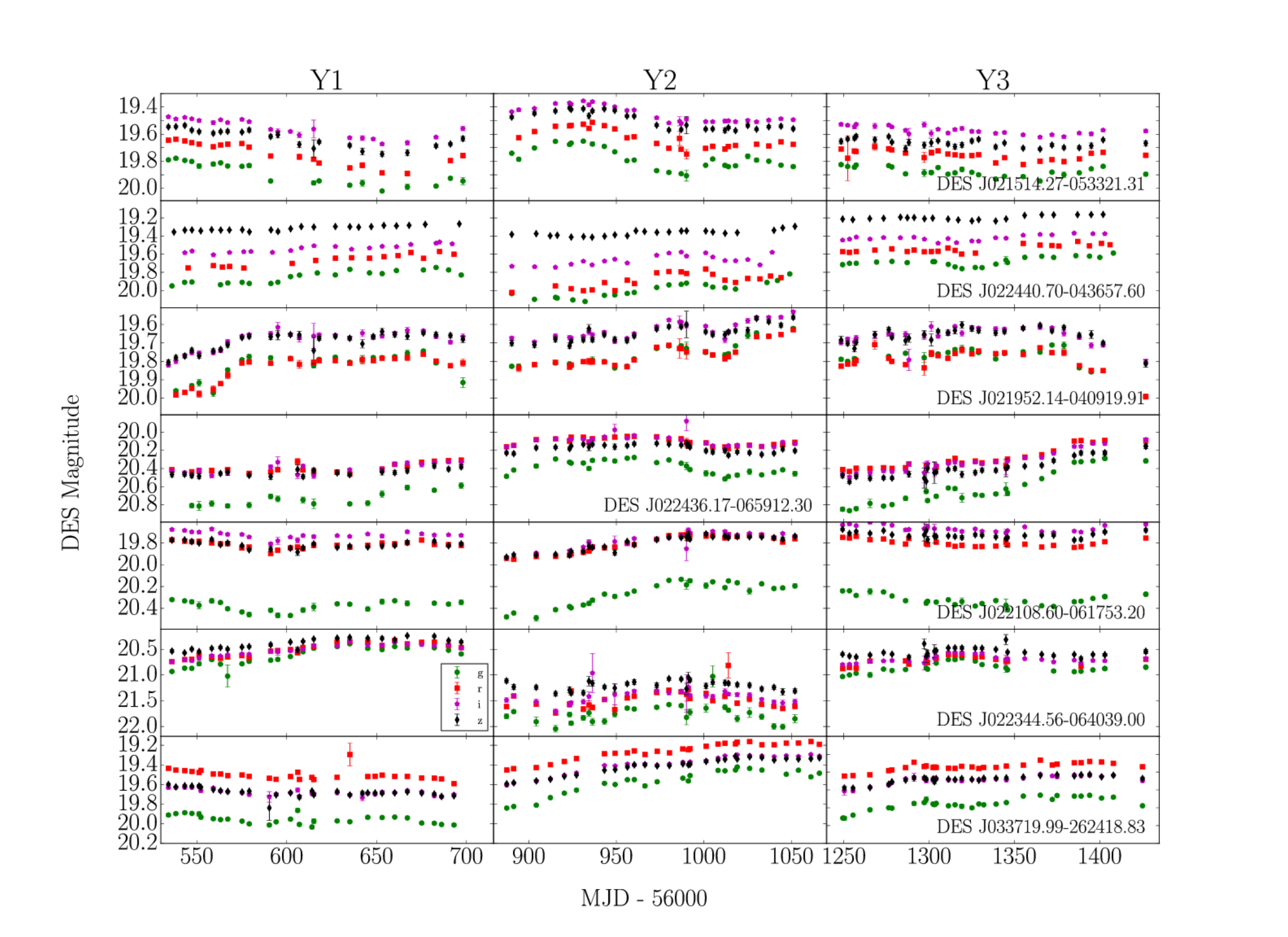}
}%
\caption{(Continued) DES quasar light curves.}
\end{figure*}

\subsection{Lag Measurements}
\label{subsec: lags}
We use two separate analysis techniques to measure the time delays between the continuum bands.  The first is the interpolated cross correlation function (ICCF) method \citep{Gaskell86, Peterson04}, where two light curves are shifted along a grid of time lags and the cross correlation coefficient $r$ is calculated at each spacing (see Figure 1 of \citealp{Gaskell86}).  This method linearly interpolates between epochs to fill in missing data.  A series of 1,000 Monte Carlo runs re-sampling (with replacement) the light curves provide the uncertainty on the lag detection using the flux randomization/random subset replacement method, where the lag distribution is given by the mean of the distribution of cross-correlation centroids for which $r > 0.8r_{\textrm{max}}$.  

For the second method, we used JAVELIN\footnote{Available at \url{https://bitbucket.org/nye17/javelin}.}, which models quasar variability as a damped random walk (DRW, \citealp{Zu13}).  The DRW models quasar light curve behavior quite well on time scales of months to years \citep{Gaskell87, MacLeod10, Zu13}, although there may be extra variability power on the much shorter timescales (tens of minutes compared to days) as seen in a group of quasars sampled by \emph{Kepler} \citep{Edelson14, Kasliwal15}.  JAVELIN has been used in previous emission line (e.g., \citealp{Zu13, Grier12b, Grier12a, Pei17}) and continuum \citep{Shappee15, Fausnaugh16} reverberation mapping campaigns.  

JAVELIN assumes that all light curves are shifted, scaled, and smoothed versions of the driving light curve.  It uses the DRW model to carry out the interpolation between epochs.  For this purpose, it is not essential that the DRW model exactly describe the true variability of the quasar - it need only be a reasonable approximation.  The model has five parameters if fitting two light curves, a driver and shifted version: the DRW amplitude and timescale, the relative flux scale factor, the top hat smoothing time scale, and the time lag.  For relatively short, sparsely-sampled light curves, it is not possible, either statistically or physically, to determine all of these parameters.  Therefore we restrict the damping timescale to 100-300 days, as has been found for a larger sample of quasars from the Sloan Digital Sky Survey (SDSS; \citealp{MacLeod10}).

\begin{figure}
  \centerline{
    \includegraphics[width=3.8in]{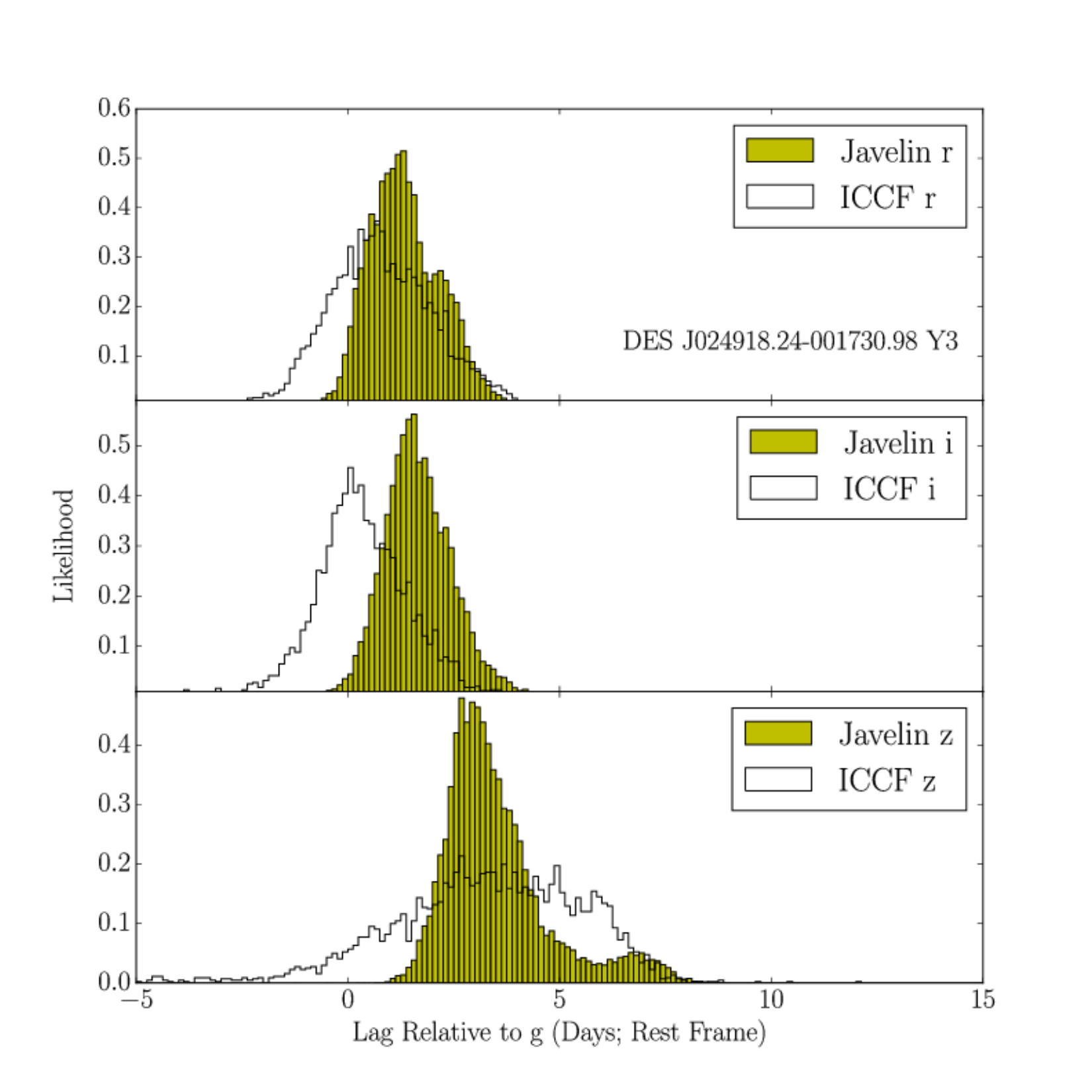}
}
  \caption{Comparison of the JAVELIN and $r > 0.8r_{\textrm{max}}$ cross-correlation centroids lag distribution for DES J024918.24-001730.98 relative to the \emph{g}-band light curve.  The two distributions are consistent with one another, although JAVELIN is much more centrally peaked than the ICCF in the \emph{z}-band distribution.  The lag distributions are given in the rest frame of the quasar.}
  \label{fig: Jav2CCF}
\end{figure}

JAVELIN also assumes that the measurement errors are well-characterized and Gaussian, this can lead to underestimation of the uncertainties if either of these assumptions is incorrect.  This has been noted in other studies (e.g., \citealp{Fausnaugh16}), and we compare our lag distributions from JAVELIN to those obtained through the ICCF in Figure \ref{fig: Jav2CCF}.  The general trend is that the two methods are consistent with one another, with the centroids of the distributions for which $r > 0.8r_{\textrm{max}}$ for the ICCF uncertainties generally being larger.  This has been noted in other works (e.g., \citealp{Fausnaugh16}).  \citet{Jiang17} perform several tests on JAVELIN with one of their quasars by creating mock light curves with a known $\delta$-function shift from the original Pan-STARRS data and find that JAVELIN is able to recover the input delays below the cadence of the light curves whereas the ICCF method does not.  We report JAVELIN lags for the remainder of the paper.

We provide a summary of our lag posterior distributions in Table \ref{tab: lag_table}.  Most of these lags are consistent with no time delay in the continuum emission at the 1-2$\sigma$ level, but the wavelength-dependent offsets in many cases strongly suggest an upper limit on the lag has been observed.  We use these upper limits, based on the 2$\sigma$ positive tail of the lag distributions, in all following analysis that use the individual JAVELIN lag data.

\subsection{JAVELIN Thin Disk Model}
\label{subsec: jav}
We created an extension to JAVELIN, hereafter the JAVELIN thin disk model, that fits directly for the parameters of the thin disk model, $R_{\lambda_{0}}$ and $\beta$, rather than getting these from an optimized fit to the individual time lags.  This Thin Disk model is now available for public use through the regular JAVELIN distribution.  We note that this is the flux-weighted radius $R_{\lambda_{0}}$ from Equation \ref{eqn: thin_disk_Fausnaugh}.  To achieve this, we adapted JAVELIN to fit all the continuum light curves simultaneously and find which $R_{\lambda_{0}}$ and $\beta$ values best reproduce the lagged light curves from the driving light curve.  The benefit of this approach is that it reduces the number of parameters and uses all the photometric data to essentially produce a better sampled light curve.  In the case where $\beta = 4/3$, the $R_{\lambda_{0}}$ parameter from Equation \ref{eq: thin_disk_short} or \ref{eqn: thin_disk_Fausnaugh} sets the absolute size of the disk and depends on the quasar properties (mass, mass accretion rate, radiative efficiency, etc.) while $\beta$ corresponds to the temperature profile of the disk as a function of radius.  If $\beta \neq 4/3$, this means that there is some heating term that does not scale as $1/R$.

\begin{figure}
  \centerline{
    \includegraphics[width=3.8in]{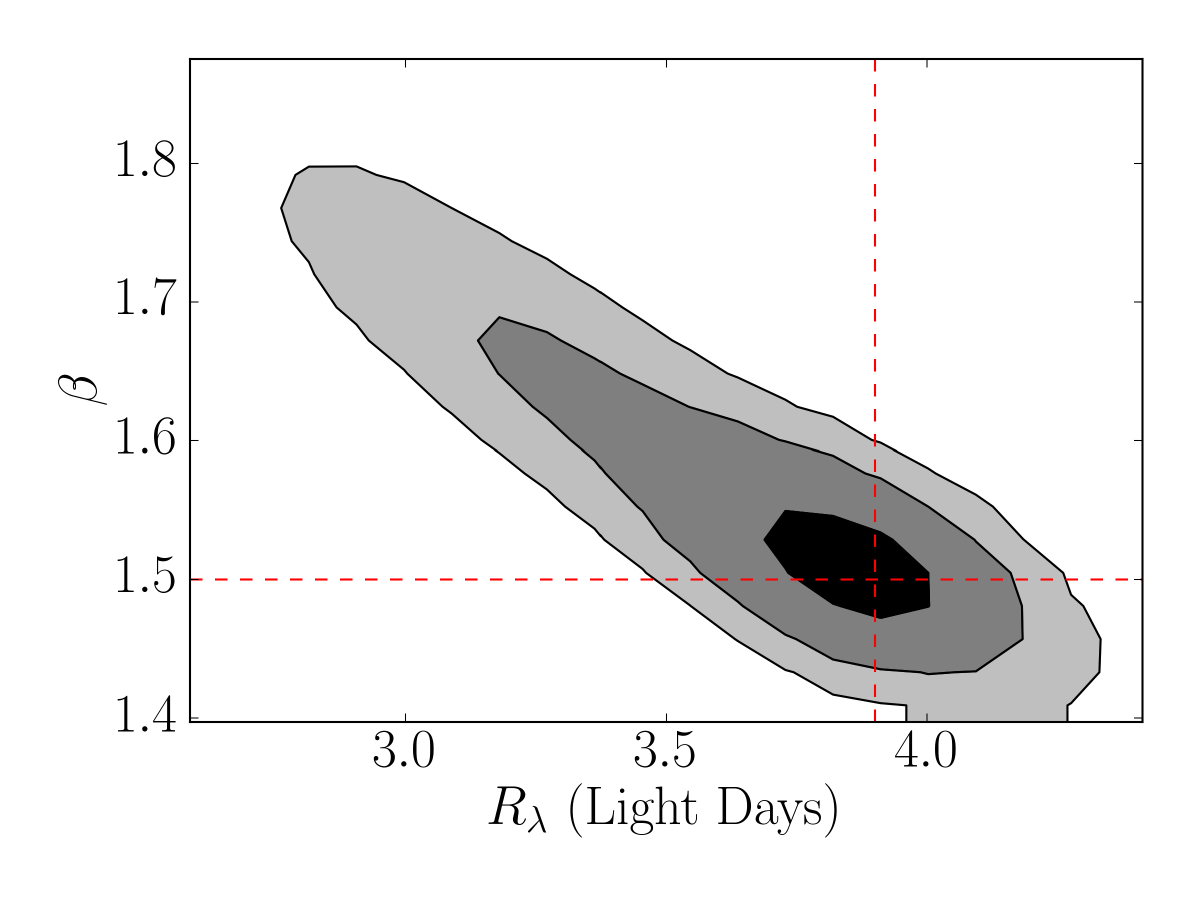}
}
  \caption{A test of our JAVELIN thin disk model's ability to recover input $R_{\lambda_{0}}$ and $\beta$ values from simulated light curves.  The dashed lines correspond to the input values used to created the lagged light curves.  Black, dark gray, and light gray correspond to values that fall within the 1, 2, and 3$\sigma$ posterior likelihood distributions.  The cadence for these data is roughly 8 days, similar to the cadence we have with DES, and we successfully recover a disk size of light travel time that is comparable to half the simulated cadence.}
  \label{fig: sim_testing}
\end{figure}

\begin{figure}
  \centerline{
    \includegraphics[width=3.8in]{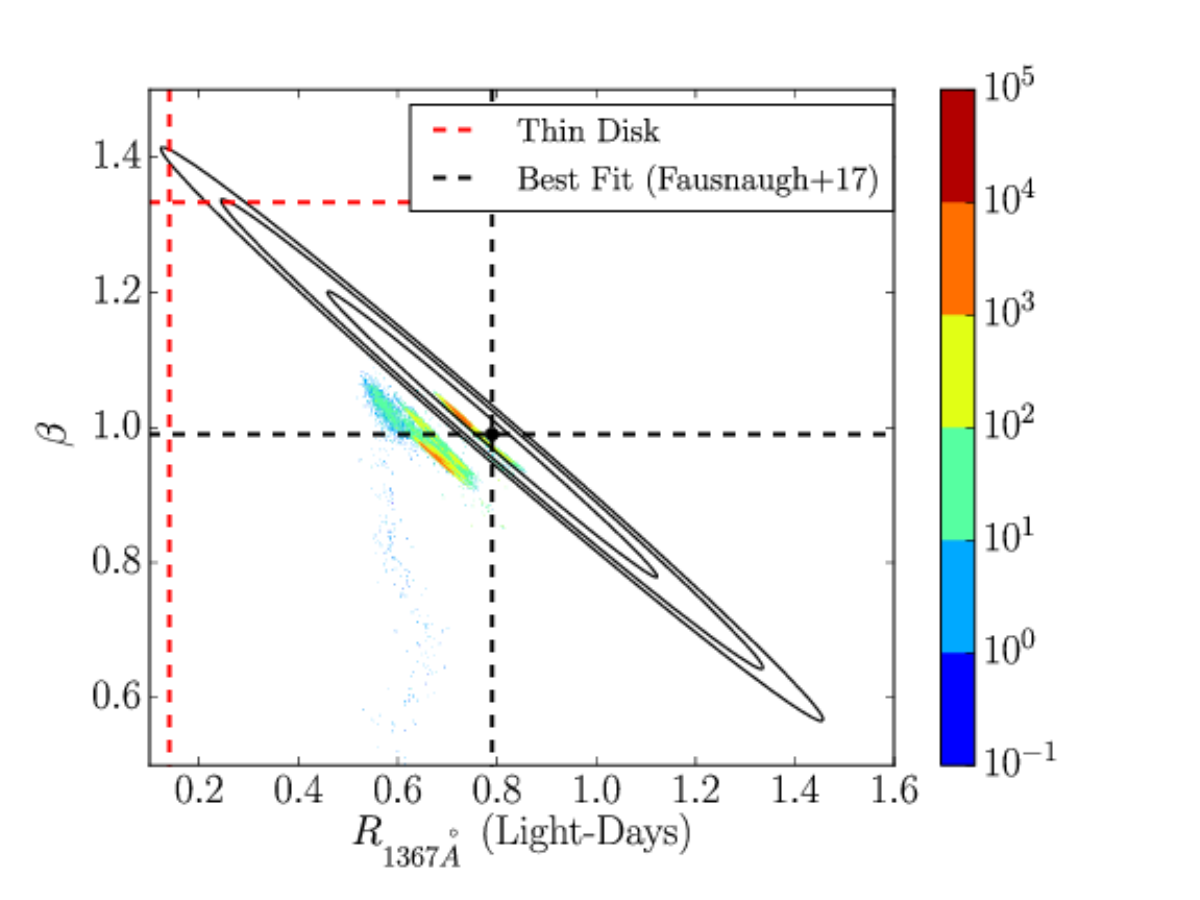}
}
  \caption{Comparison of the accretion disk size at 1367\AA\enskip for NGC 5548 using our new JAVELIN thin disk model (colored points) versus the results reported in (\citealp{Fausnaugh16}; black contours).  The black contours correspond to their 1, 2, and 3$\sigma$ constraints on these accretion disk parameters based on a fit to the lags measured relative to 1357\AA, whereas the colored points represent the relative posterior MCMC probability distribution for these same parameters.  We find two families of solutions of approximately equal likelihood, one of which is within the 1$\sigma$ contours of the previous NGC 5548 analysis.  We are consistent in finding a disk profile that is slightly shallower and disk size much larger than predicted based on the mass and luminosity of NGC 5548, which is given by the intersection of the red dashed lines.}
  \label{fig: 5548_alphabeta}
\end{figure}

We tested these modifications to JAVELIN in two ways.  First, the JAVELIN website provides a simulated 5 year quasar light curve, with 8 day cadence and 180 day seasonal gaps to account for realistic seasonable inaccessibility, similar to what we will expect at the end of DES.  We used a grid of $R_{\lambda_{0}}$ and $\beta$ values to generate new light curves based on these simulated data with known lags between the bands.  We then analyzed these simulated light curves to determine how well these values were recovered in several regimes.  Most critically, we were interested in the performance when the light curve sampling rate was greater than, comparable to, and less than the shifts we applied using our known input thin disk parameters for $R_{\lambda_{0}}$ and $\beta$.  Recovery of these values works well in many cases, although it has some trouble recovering the model parameters in the instances with steep temperature profiles ($\beta >$ 3) or disk light-crossing times $R_{\lambda_{0}}/c$ that are small compared to the sampling timescale ($<$1/10th of the cadence).  Figure \ref{fig: sim_testing} shows the results of a test with $R_{\lambda_{0}} = 3.9$ light-days and $\beta = 1.5$.  In this test, the parameters are recovered for a temperature profile close to what is predicted by thin disk theory, and with a disk size of the driving light curve at slightly under half the sampling cadence of the data.

\begin{figure}
  \centerline{
    \includegraphics[width=3.8in]{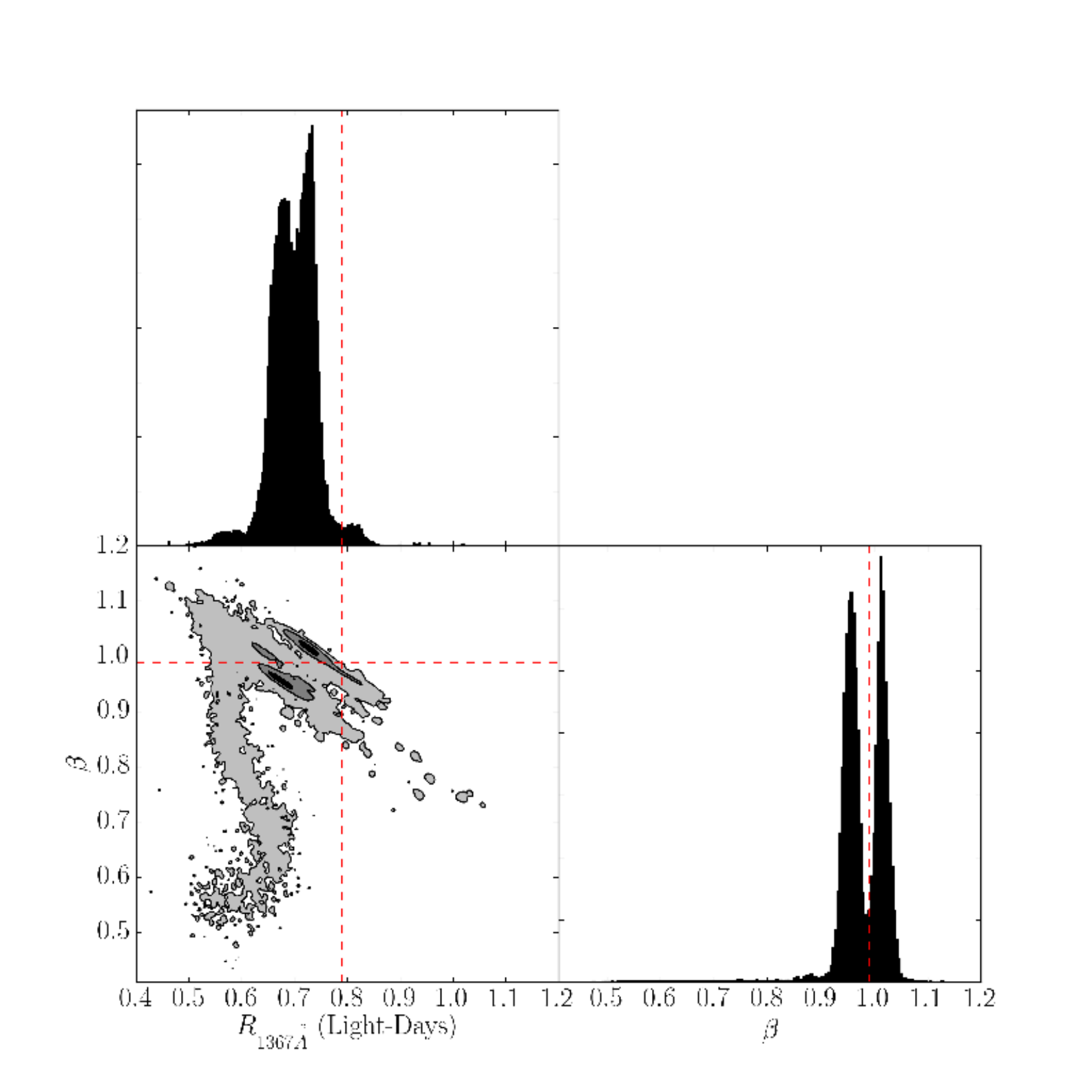}
}
  \caption{Corner plot for the $R_{\lambda_{0}}$ and $\beta$ parameters of NGC 5548 using our JAVELIN thin disk model.  These histograms show clearly the double-valued nature of the fits presented in Figure \ref{fig: 5548_alphabeta}.  The top and right panels show histograms for $R_{\lambda_{0}}$ and $\beta$ individually, and the lower left panel illustrates the covariance between the two.  As with the previous figure, black, dark gray, and light gray correspond to values that fall within the 1, 2, and 3$\sigma$ likelihood distributions.  The dashed lines give the best fit values found in \citet{Fausnaugh16}.}
  \label{fig: 5548_triangle}
\end{figure}

We also reanalyzed the data from \citet{Fausnaugh16} on NGC 5548.  \citet{Fausnaugh16} performed pairwise lag analyses on sets of \emph{Hubble Space Telescope} (\emph{HST}), \emph{Swift}, and ground-based light curves to detect lags with respect to the HST 1367\AA\enskip light curve and then fit for $R_{\lambda_{1367\textrm{\AA}}}$ ($\alpha$ in that paper) and $\beta$.  Those authors experimented with various subsets of the data, particularly subsets that excluded bands that may be contaminated by emission from other physical processes (e.g., broad emission lines or the Balmer continuum).  We reanalyzed their data with our modified code and simultaneously fit a total of 17 light curves excluding the \emph{U} and \emph{u} bands due to Balmer continuum contamination and fixing the damping timescale at 164 days.  The \emph{u}- and \emph{U}-band exclusion was adopted by \citet{Fausnaugh16} and the damping timescale is based on previous, longer time baseline studies of NGC 5548 by \citet{Zu11}.  Our result on the NGC 5548 data is consistent with the \citet{Fausnaugh16} values with tighter constraints on $R_{\lambda_{1367\textrm{\AA}}}$ and $\beta$ from fitting them directly, albeit now with a double solution, shown in Figures \ref{fig: 5548_alphabeta} and \ref{fig: 5548_triangle}.  The two solutions have similar likelihood to one another, and prefer smaller $R_{\lambda_{1367\textrm{\AA}}}$ and $\beta$ values compared to those found in \citet{Fausnaugh16}.  One of the two solutions is within the 1$\sigma$ error contours from \citet{Fausnaugh16}.  The double-peaked nature of our solution may be a product of the high dimensionality of parameter space in our model, and it is encouraging that we find a similar result as a different method.

\subsubsection{DES Results}
\label{subsubsec: DES_res}
\begin{figure}
 \centerline{
   \includegraphics[width=3.8in]{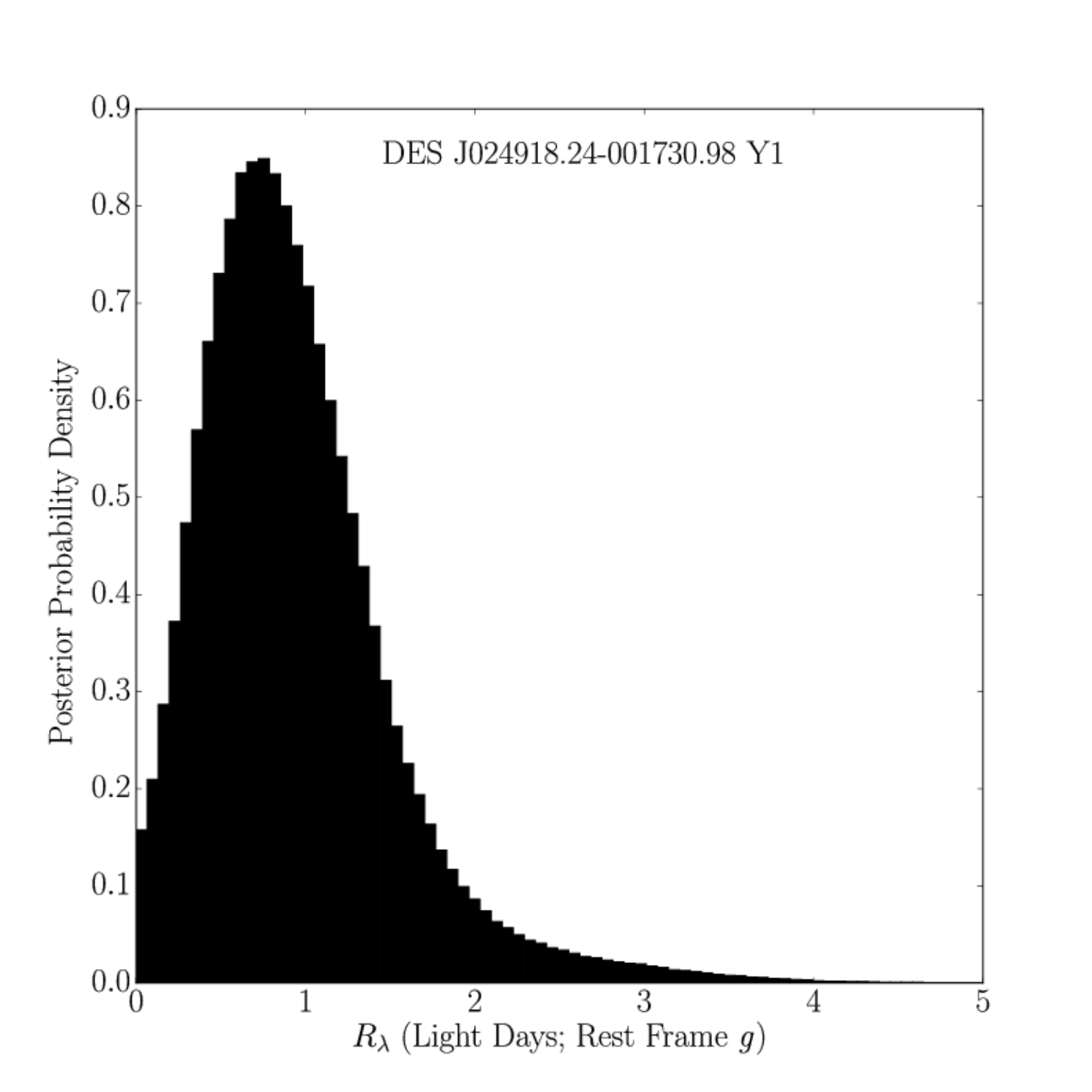}
 }
 \caption{Example $R_{\lambda_{0}}$ posterior distribution after fitting with our new JAVELIN thin disk model and fixing $\beta = 4/3$ for DES J024918.24-001730.98, the same object from Figure \ref{fig: Jav2CCF}.  The $x$-axis is the quasar accretion disk size (in light-days) at the wavelength corresponding to the emitting region of the observed DES \emph{g} band, which runs from approximately 2000-3000\AA\textrm{} rest-frame for our sample given its redshift distribution.}
 \label{fig: alpha_post}
\end{figure}

We then used this algorithm for the DES quasars that had yielded tentative \emph{r}, \emph{i}, and/or \emph{z} lag measurements relative to \emph{g} band.  We still restrict the DRW damping time scale to fall between 100-300 days for this analysis.  We also fixed $\beta$ = 4/3, as we found we were unable to provide good constraints on both $R_{\lambda_{0}}$ and $\beta$ simultaneously for any object.  This assumption is well-motivated from theory, but we note that observations have shown evidence for a smaller $\beta$ value closer to 0.8-1.  If $\beta$ is smaller than our assumption, this would inflate the final disk sizes somewhat.  Figure \ref{fig: alpha_post} shows an example posterior distribution for $R_{\lambda_{0}}$ for DES J024918.24-001730.98 whose full posterior distributions for all model parameters is in Figure \ref{fig: full_corner_example}.  Fits for the remainder of our objects can be found in the Appendix.

As stated in Section \ref{sec: analysis}, five objects in our current quasar sample have large amplitude variability in more than one DES observing season, and we compare their disk sizes with this method to check its performance.  Two of these objects have disk sizes from three DES seasons, while the other three have only two analyzed seasons of variability.  Figure \ref{fig: multiple_years} illustrates the $R_{\lambda_{0}}$ values for each of these objects, and we see that they are all consistent at the 1-1.5$\sigma$ level between the observing seasons.  Different disk sizes between seasons could imply that the size uncertainties are underestimated, that the disk has undergone some structural change, or that the time delays are being influenced by emission that is not a simple reprocessing of emission from the inner regions of the disk.

\begin{figure*}
  \centerline{
    \includegraphics[width=8in]{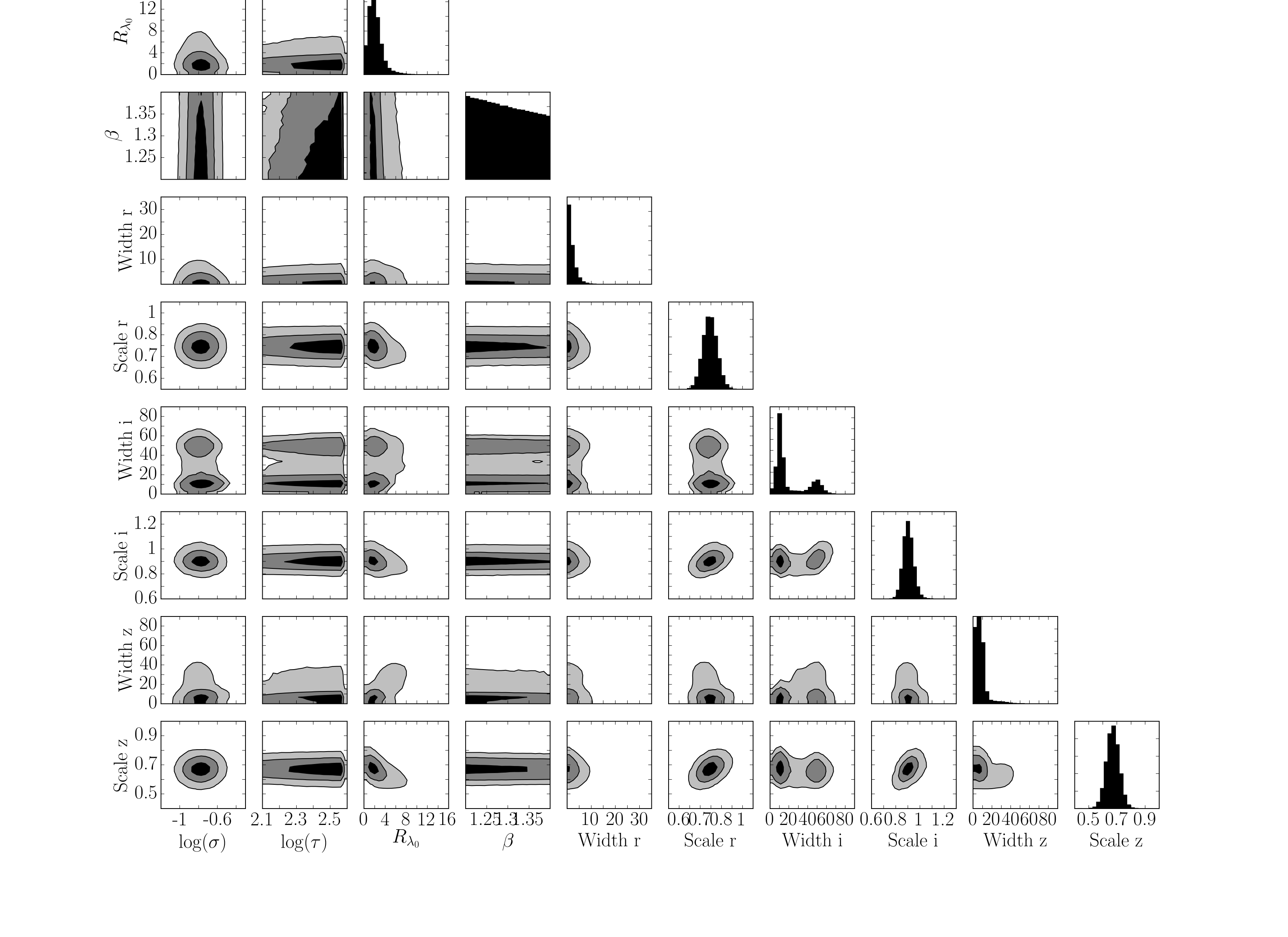}
}
  \caption{Corner plot for the full parameters for the quasar presented in Figure \ref{fig: Jav2CCF}, DES J024918.24-001730.98 (Y1), in the JAVELIN thin disk model analysis.  As mentioned in the text, strong priors were placed on both the damping timescale $\tau$ and wavelength dependence $\beta$.  The contours progress from 1-3$\sigma$ as they transition from black to lighter grey.  The $R_{\lambda_{0}}$ parameter is in units of light-days, whereas $\tau$ and all of the widths are in days.  Note that all parameters are in the observed frame.}
  \label{fig: full_corner_example}
\end{figure*}

\begin{deluxetable}{lrrrrr}
\tabletypesize{\tiny}
\tablewidth{3.5truein}
\tablecaption{$R_{2500\textrm{\AA}}$ and Lags}
\tablehead{
\colhead{Object Name} &
\colhead{$\tau_{r}$} &
\colhead{$\tau_{i}$} &
\colhead{$\tau_{z}$} &
\colhead{$R_{2500\textrm{\AA}}^{1}$} &
\colhead{$R_{2500\textrm{\AA}}^{2}$} \\
\colhead{(Season)} &
\colhead{Days} &
\colhead{Days} &
\colhead{Days} &
\colhead{Lt-Days} &
\colhead{Lt-Days} 
}

\startdata
DES J0224-0436 (Y1) & 2.3$_{-11.1}^{+7.8}$ & 2.9$_{-4.7}^{+4.5}$ & 3.4$_{-5.4}^{+4.5}$ & 8.6 & 2.3$_{-0.8}^{+1.6}$ \\ 
DES J0243-0113 (Y1) & 0.1$_{-1.9}^{+1.7}$ & 0.8$_{-1.9}^{+2.2}$ & 1.2$_{-2.3}^{+2.2}$ & 1.8 & 0.8$_{-0.8}^{+1.5}$ \\ 
DES J0253+0004 (Y1) & 2.0$_{-2.7}^{+5.4}$ & 2.3$_{-7.2}^{+7.6}$ & 5.0$_{-5.4}^{+7.6}$ & 10.8 & 7.3$_{-5.4}^{+5.8}$ \\
DES J0253+0004 (Y2) & 1.0$_{-1.8}^{+2.5}$ & 0.9$_{-1.8}^{+2.5}$ & 1.3$_{-2.1}^{+2.5}$ & 4.4 & 1.3$_{-1.6}^{+1.4}$ \\ 
DES J0249-0017 (Y1) & 1.0$_{-1.1}^{+1.2}$ & 0.4$_{-0.9}^{+5.5}$ & 1.0$_{-1.2}^{+5.5}$ & 2.4 & 0.9$_{-0.7}^{+1.6}$ \\ 
DES J0249-0017 (Y3) & 1.3$_{-1.3}^{+1.8}$ & 1.6$_{-1.4}^{+1.9}$ & 3.2$_{-1.4}^{+1.9}$ & 4.5 & 3.0$_{-1.2}^{+1.8}$ \\ 
DES J0224-0659 (Y1) & 1.2$_{-1.3}^{+1.8}$ & 1.5$_{-1.5}^{+2.7}$ & 2.2$_{-1.9}^{+2.7}$ & 4.3 & 7.8$_{-6.4}^{+5.7}$ \\
DES J0224-0659 (Y2) & 0.5$_{-1.2}^{+1.1}$ & 0.6$_{-1.1}^{+1.6}$ & 2.1$_{-1.5}^{+1.6}$ & 2.6 & 2.4$_{-1.4}^{+1.6}$ \\ 
DES J0224-0659 (Y3) & 3.3$_{-4.9}^{+3.5}$ & -1.3$_{-4.5}^{+4.1}$ & 4.9$_{-11.5}^{+4.1}$ & 8.3 & 5.3$_{-3.0}^{+3.0}$ \\ 
DES J0221-0617 (Y1) & 1.6$_{-2.4}^{+5.4}$ & 4.7$_{-5.6}^{+3.5}$ & 3.9$_{-5.0}^{+3.5}$ & 8.6 & 2.7$_{-2.1}^{+6.8}$ \\ 
DES J0221-0617 (Y2) & 2.4$_{-3.9}^{+2.2}$ & 3.4$_{-4.5}^{+2.1}$ & 2.3$_{-2.0}^{+2.1}$ & 5.9 & 3.1$_{-2.7}^{+8.5}$ \\ 
DES J0221-0617 (Y3) & 0.1$_{-2.1}^{+8.3}$ & 0.9$_{-1.9}^{+8.8}$ & -0.6$_{-1.8}^{+8.8}$ & 6.6 & 1.0$_{-0.9}^{+1.7}$ \\ 
DES J0219-0409 (Y2) & 0.6$_{-0.9}^{+1.1}$ & 0.7$_{-1.0}^{+0.9}$ & 0.7$_{-1.1}^{+0.9}$ & 1.4 & 0.5$_{-0.6}^{+1.0}$ \\ 
DES J0248+0010 (Y1) & 1.5$_{-1.9}^{+1.0}$ & 1.7$_{-1.1}^{+0.6}$ & 1.5$_{-2.5}^{+0.6}$ & 3.1 & 1.1$_{-0.7}^{+0.8}$ \\ 
DES J0223-0640 (Y1) & 0.5$_{-1.8}^{+1.4}$ & 0.9$_{-1.8}^{+1.4}$ & 1.4$_{-2.2}^{+1.4}$ & 2.1 & 0.9$_{-0.9}^{+8.9}$ \\ 
DES J0215-0533 (Y1) & 1.2$_{-2.7}^{+1.6}$ & 1.4$_{-2.8}^{+1.9}$ & 2.6$_{-3.7}^{+1.9}$ & 2.7 & 1.2$_{-1.4}^{+2.0}$ \\ 
DES J0215-0533 (Y2) & 3.2$_{-1.7}^{+3.3}$ & 3.7$_{-2.2}^{+3.1}$ & 3.8$_{-0.7}^{+3.1}$ & 5.7 & 2.6$_{-1.7}^{+1.6}$ \\ 
DES J0241-0105 (Y3) & 1.4$_{-2.5}^{+2.7}$ & 2.3$_{-3.5}^{+3.0}$ & 3.3$_{-3.6}^{+3.0}$ & 4.5 & 2.2$_{-2.0}^{+2.2}$ \\ 
DES J0241-0105 (Y2) & 0.8$_{-1.7}^{+1.7}$ & 1.0$_{-1.5}^{+2.2}$ & -0.2$_{-2.3}^{+2.2}$ & 1.8 & 1.2$_{-1.2}^{+1.4}$ \\ 
DES J0241-0107 (Y1) & 0.0$_{-1.1}^{+1.8}$ & 0.9$_{-1.5}^{+1.7}$ & 1.9$_{-2.5}^{+1.7}$ & 3.6 & 2.9$_{-1.5}^{+2.4}$ \\ 
DES J0247-0021 (Y1) & 0.7$_{-1.6}^{+2.5}$ & 1.0$_{-2.5}^{+1.7}$ & 1.7$_{-2.6}^{+1.7}$ & 4.1 & 1.7$_{-0.9}^{+1.7}$ \\ 
DES J0247-0021 (Y2) & -1.1$_{-1.6}^{+2.9}$ & -0.2$_{-2.7}^{+2.3}$ & 0.5$_{-2.8}^{+2.3}$ & 1.7 & 1.7$_{-1.3}^{+1.6}$ \\ 
DES J0215-0430 (Y2) & 2.9$_{-2.0}^{+3.8}$ & 3.4$_{-4.7}^{+2.8}$ & 3.3$_{-6.3}^{+2.8}$ & 7.0 & 7.6$_{-7.5}^{+6.4}$ \\ 
DES J0337-2624 (Y2) & 1.3$_{-1.0}^{+1.0}$ & 2.0$_{-1.4}^{+0.9}$ & 2.4$_{-1.9}^{+0.9}$ & 3.5 & 3.4$_{-1.7}^{+1.7}$ \\ 
\enddata
\label{tab: lag_table}
\tablecomments{JAVELIN time lags and disk sizes ($eta$=4/3) in the quasar rest frame.  $^{1}$Disk size 2$\sigma$ upper limits from \emph{riz} lags.  $^{2}$Disk sizes from the new thin disk model with 2$\sigma$ error bars.  Note that these are flux-weighted radii from Equation \ref{eqn: thin_disk_Fausnaugh}.}
\end{deluxetable}

\begin{figure}
  \centerline{
    \includegraphics[width=3.8in]{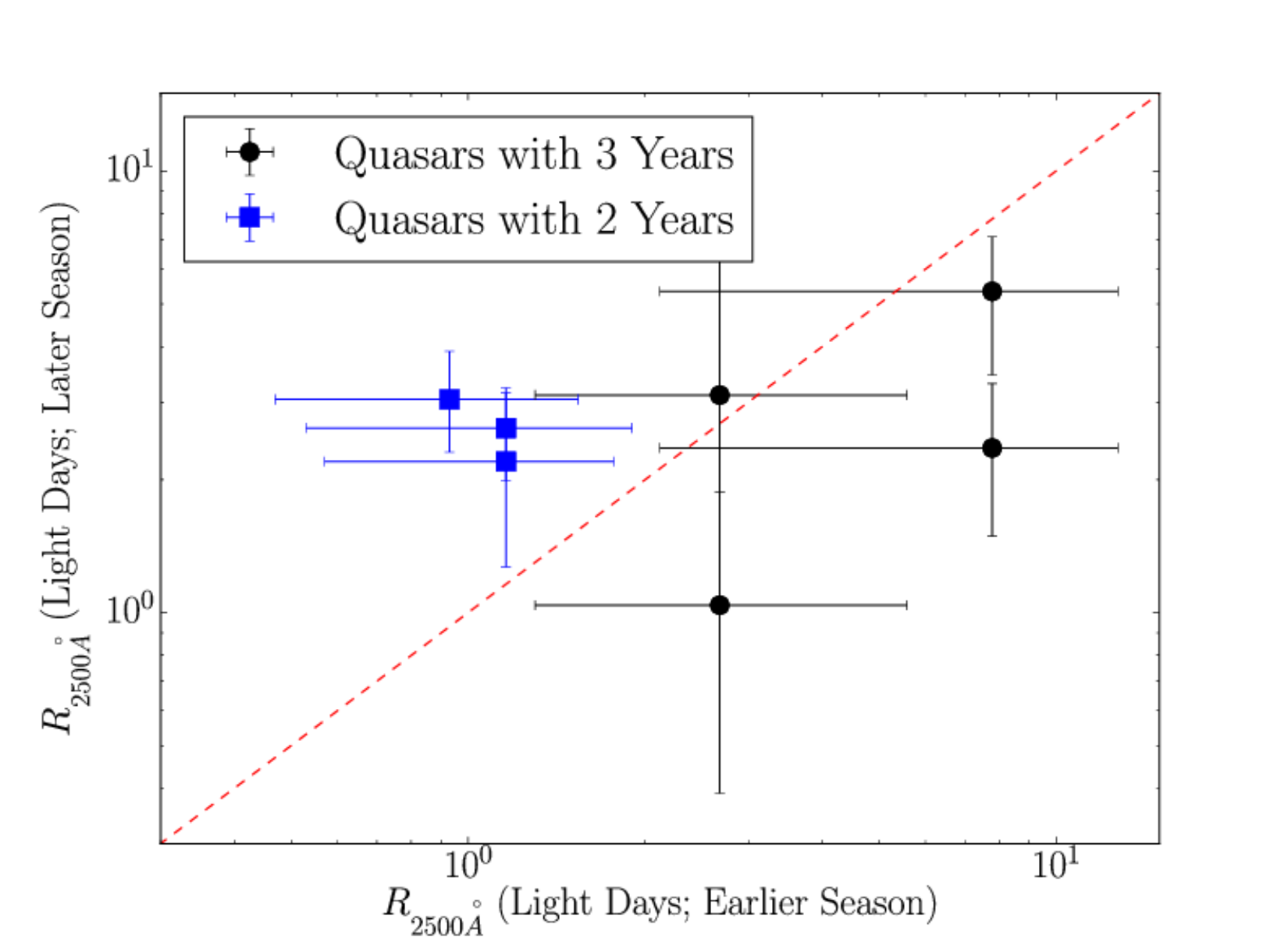}
}
  \caption{Comparison of the JAVELIN disk model results for objects that have detections in multiple years.  The three square blue points are the three quasars that have strong enough variability in two DES seasons to get two separate $R_{\lambda_{0}}$ measurements, whereas the black points are measurements for two quasars that have variability in all three seasons.  Both of these quasars thus have two points on the plot, one comparing year 1 to year 2, and then another comparing year 1 to year 3.  The diagonal dashed line is a 1:1 relation.  The error bars represent the $1\sigma$ limits on the parameter distributions, and show that the multiple years are roughly consistent with one another.}
  \label{fig: multiple_years}
\end{figure}

One possible concern is contamination of the photometric bandpasses by emission line flux from larger scales. For the redshift range of our sources DES (0.7 $<$ z $<$ 1.6), the \mgii\enskip emission line is present in the \emph{g}, \emph{r}, or \emph{i} band, along with surrounding blends of Fe emission.  The equivalent widths of the \civ\enskip line for luminous quasars like those in our sample are lower than for less luminous Seyferts like NGC 5548 (the well-known Baldwin Effect; \citealp{Baldwin77}).  We therefore expect emission line contamination to be less important for our quasars that have \civ\enskip emission within the DES bandpass (a small fraction of our sample) compared to AGN like NGC 5548.  For the sample of Pan-STARRS quasars, similarly high luminosity to those presented here, \citet{Jiang17} find that the emission line contamination to continuum flux for their highest confidence subsample is less than 6\% and make a negligible contribution to the lag signals.  Using the OzDES coadded spectra, we find that the \mgii\enskip line contribution to the total flux in the DES bandpasses is roughly the same as was found in \citet{Jiang17}, about 1-3\% with a maximum of 7\%.  What matters for lag measurements for the accretion disk is that the strength of any line variability in the bandpass is much smaller than that of the continuum on short timescales.  The line to continuum flux ratio is a reasonable proxy for this effect, although the relative line variability is often larger than the relative continuum variability.

\subsection{Correlation with Black Hole Masses}
\label{subsec: masses}
To compare our disk sizes with previous studies, we require an estimate of the black hole mass.  A reverberation mapping campaign is currently underway for these sources (see \citealp{King15} for a simulation of the DES reverberation mapping survey) which will provide the most accurate masses.  For now, we use single epoch mass estimates using the \mgii\enskip broad emission line from the OzDES coadded spectra and the relationship from \citet{McLure02}.  The OzDES spectra were calibrated using the DES photometry, and the emission line full-width half-maxima (FWHM) were measured using the IRAF \emph{splot} package.  The resulting masses span more than an order of magnitude.

\begin{figure}
  \centerline{
    \includegraphics[width=3.8in]{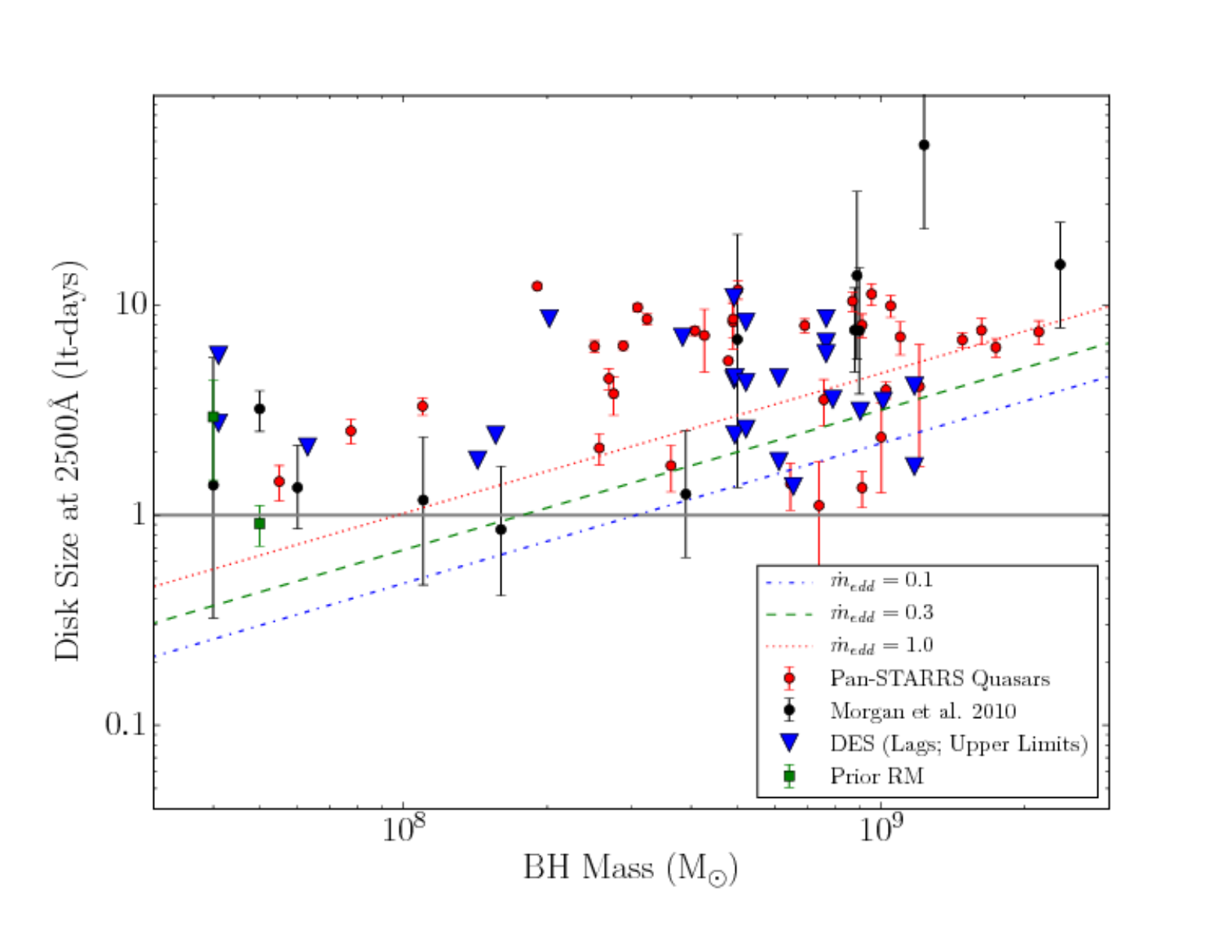}
}
  \caption{Accretion disk size distribution at 2500\AA\textrm{} as a function of black hole mass.  The two slanted lines correspond to the predictions from thin disk theory for an accretion disk inflated by a factor of $X^{4/3}$ per the prescription in \citet{Fausnaugh16} and found in Equation \ref{eqn: thin_disk_Fausnaugh} at three different accretion rates relative to Eddington- 0.1, 0.3, and 1.  The DES sample is shown as upper limits based on the 95th percentile of the JAVELIN lag distributions as a conservative estimate given that several of our \emph{g}-\emph{r} and \emph{g}-\emph{i} lags were 2$\sigma$ consistent with no lag.  The black points are taken from gravitational lensing measurements of quasar accretion disk sizes \citep{Morgan10}.  As these were reported in units of the radius where $\lambda = \frac{hc}{kT}$, they are similarly inflated to match the flux-weighted radius that is measured by the time-delay reverberation mapping data.  The green points are the result of photometric reverberation mapping of NGC 5548 from the STORM campaign \citep{Edelson15, Fausnaugh16} and NGC 2617 \citep{Shappee15}, and the magenta for Pan-STARRS \citep{Jiang17}.  Note that we assumed that the disk's size scaling as a function of wavelength, $\beta$, is 4/3 to put all the objects at the same rest wavelength.}
  \label{fig: massplot_ind}
\end{figure}

We compare our results to other studies in Figure \ref{fig: massplot_ind}.  To meaningfully compare all objects at the same rest wavelength of 2500\AA, we assume that the effective accretion disk size scales as the predicted $\beta = 4/3$.  Since the microlensing models output the radius where $\lambda = \frac{hc}{k_{B}T}$ and our photometric data are sensitive to the flux-weighted radius that we have parameterized as $\lambda = \frac{Xhc}{k_{B}T}$ in Equation \ref{eqn: thin_disk_Fausnaugh}, we inflate the microlensing results by $X^{4/3}$, calculated from Equation \ref{eqn: X} for each object individually, such that all of the quasar radii between the two methods are probing the same disk scale.  Despite presenting our redshift-corrected individual lags results as upper limits derived from the 2$\sigma$ positive tail of their lag distributions because the negative tails are consistent with no lag at the 1$\sigma$ level, they are fairly similar to measurements reported by other studies.  

\begin{figure}
  \centerline{
    \includegraphics[width=3.8in]{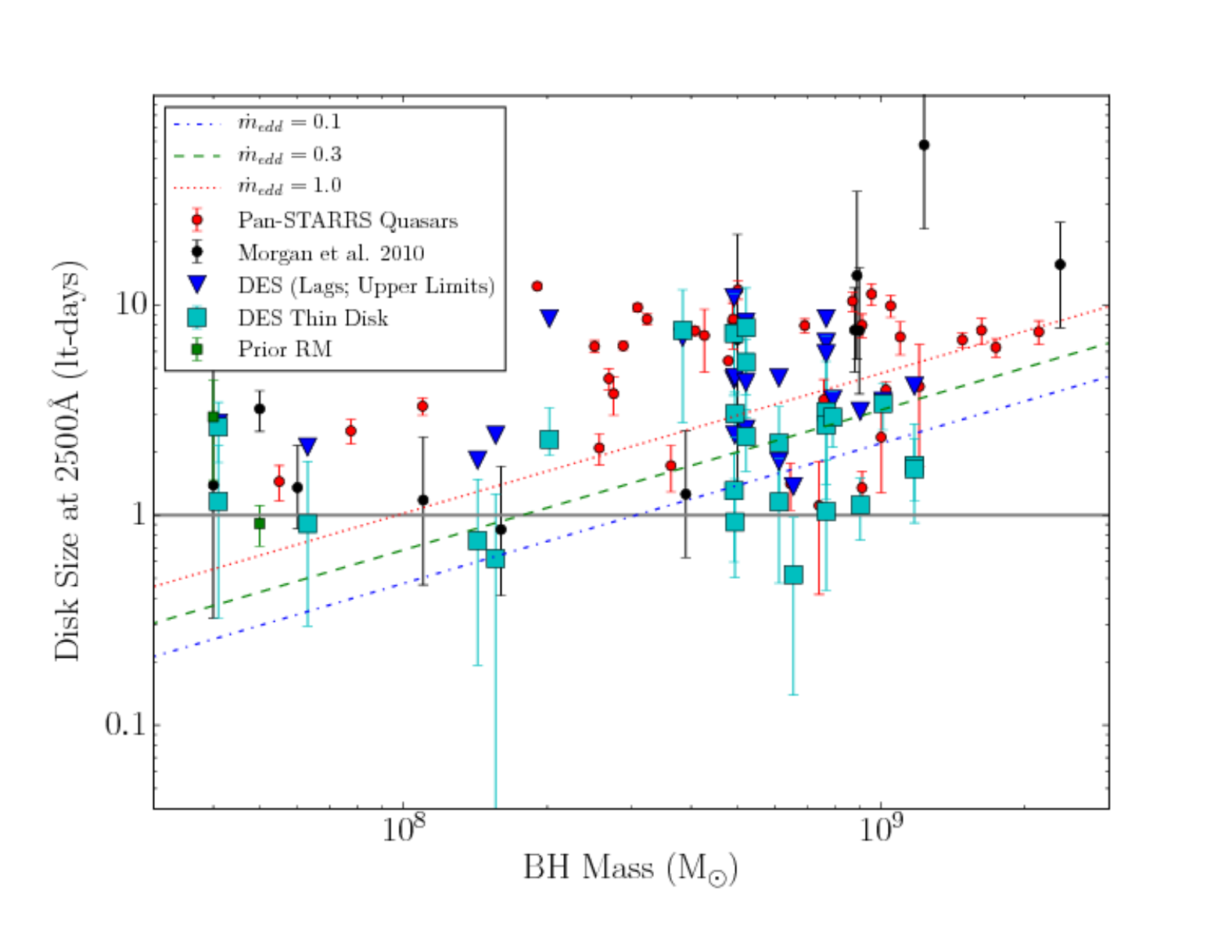}
}
  \caption{Disk size as a function of black hole mass, similar to Figure \ref{fig: massplot_ind}, but now with our disk sizes obtained from our new Thin Disk JAVELIN extension as cyan squares.  As before, all objects have been scaled to the flux-weighted radius in Equation \ref{eqn: thin_disk_Fausnaugh} rather than the radius where $\lambda = \frac{hc}{k_{B}T}$.}
  \label{fig: mass_plot_new_jav}
\end{figure}

We then compared these $R_{\lambda_{0}}$ constraints from our JAVELIN model to those derived from fitting the lags alone.  Figure \ref{fig: mass_plot_new_jav} shows these new size measurements alongside both literature values and those shown in Figure \ref{fig: massplot_ind}, and we provide a summary of these disk sizes in Table \ref{tab: lag_table}.  Many of our quasar disk sizes are consistent with moderate ($\sim$0.3) to super-Eddington accretion rates as well as the larger disk sizes of the previous literature given our uncertainties.  While many of our measurements are on the lower end of previous disk sizes, the DES sample spans the same range of disk sizes as those found in the Pan-STARRS survey \citep{Jiang17}.  Both sets of disk sizes assume a temperature profile $\beta = 4/3$.  The principle difference is that our method assumes a thin disk structure from the outset and fits all light curves simultaneously.  We sample the thin disk parameters and use them to produce lagged light curves which are matched to the observed light curves and then evaluated on how well they reproduce the observations.  This method does not fit for lags directly, instead they are produced from Equation \ref{eq: thin_disk_lag} given our observed wavelengths and MCMC sampled model parameters.  Previous works have instead fit for the lags first, pairwise with respect to the continuum, and then used the ensemble of lags with respect to the continuum to fit for a disk size.  There is no immediately obvious reason why one method would produce systematically larger or smaller accretion disks than the other.

From Equation \ref{eqnarray: thin_disk_Kochanek}, we expect that the disk size should scale as $R_{\lambda_{0}} \propto M_{\textrm{BH}}^{2/3}$ in traditional thin disk theory for a roughly fixed $L/L_{\textrm{E}}$, as has been found for luminous quasars \citep{Kollmeier06}.  We note, however, that X-ray studies (e.g., \citealp{Aird12}), optical studies with well-characterized selection functions (e.g., \citealp{Schulze10}), and phenomenological modeling (e.g., \citealp{Weigel17}) argue that the Eddington ratio distribution of quasars is more power law-like.  There appears to be a weak trend with mass in Figure \ref{fig: mass_plot_new_jav}, albeit with a large scatter.  It should be noted again that the uncertainties in the mass measurements for all of these quasars is roughly 0.4 dex.  Prior accretion disk measurements, both through microlensing and photometric reverberation mapping, have found that many disks larger than expected from thin disk theory by a factor of a few \citep{Morgan10, Fausnaugh16, Jiang17}.  Given our large uncertainties, our objects are consistent with both these larger disk sizes as well as thin disks with moderate-to-high accretion rates.  A small subset of our JAVELIN thin disk models require Eddington ratios greater than unity, but our sample on the whole shows a large scatter around a thin disk with a moderate accretion rate.  

One explanation for larger disk sizes is higher accretion rates for the black holes closer to the Eddington limit \citep{Fausnaugh16}, but at these rates the disks would no longer remain thin.  Another explanation provided by \citet{Hall17} for large disks is that there could be a low-density atmosphere around the thin disk that modifies the optical and UV regions of the disk to appear larger than they would as a blackbody spectrum.  A third possibility is that quasars have high intrinsic reddening that has not been taken into account \citep{Gaskell04}.  This would make the quasars more luminous and naturally increase the expected disk sizes by a factor of a few.  Depending on the exact magnitude of this correction, it could completely remove the discrepancy of the previously too large disk sizes \citep{Gaskell17}, and move the DES quasars to Eddingtion accretion rates of 0.1 or less.  \citet{Gaskell04} also find that the intrinsic reddening is greater for lower luminosity objects, which would produce the largest discrepancy at the lowest luminosities.  While Figure \ref{fig: mass_plot_new_jav} plots the size versus black hole mass, the x-axis is a fair proxy for the luminosity.  The objects on the left of the plot are those that would have the largest reddening correction, and are currently the most discrepant from the prediction without it.

\subsection{Stacking Analysis}
\label{subsec: stacking}
\begin{figure}
  \centerline{
    \includegraphics[width=3.8in]{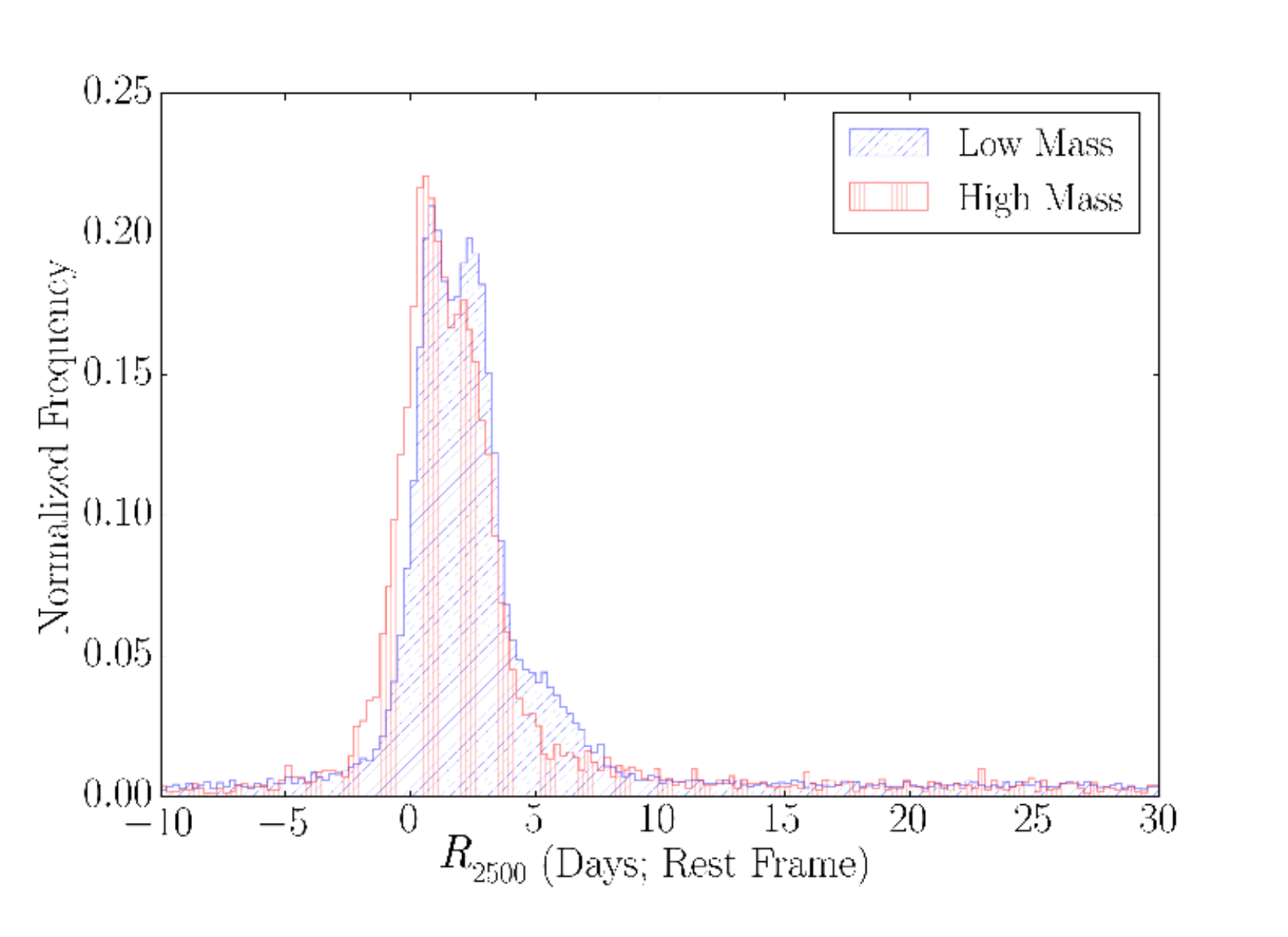}
}
  \caption{Stacked distribution of disk sizes at rest frame 2500\AA\textrm{} assuming that $\beta = 4/3$.  The two mass bins are divided at $6\times10^{7}\msol$ to give roughly equal numbers in each bin, although the dynamic range of the higher mass bin is smaller than the overall mass scaling uncertainties.  We expect a bigger disk for higher mass black holes, but the two distributions are roughly the same.}
  \label{fig: stackedjav}
\end{figure}

After analyzing the individual lag distributions, we investigated whether a stronger signal could be found by combining the posterior distributions for quasars of similar properties.  In the standard thin disk model, the absolute size of the disk depends on the quasar properties.  Thus, we expect that quasars with similar properties should have similar accretion disk sizes, and by combining their size distributions we may amplify the total signal.  We divided our sample into two bins split at a mass of $6\times10^{8}\msol$.  This value gives approximately equal numbers of objects in each bin (7 and 8), albeit with different dynamic ranges of masses.  The small mass bin covers an order of magnitude in mass, while the larger bin only a factor of two.  Given the large systematic uncertainties in the single epoch mass measurements, several of the objects could move between the bins with their mass errors, which could bias the stacked distributions.  We rerun our JAVELIN thin disk object again on each quasar individually without any priors on the parameters and then sum the accretion disk size likelihoods for all the quasars in the same bin after putting them in the rest frame and scaling them to the same reference wavelength (2500\AA) assuming $\beta = 4/3$.  Figure \ref{fig: stackedjav} shows the final distributions for our mass bins, which are consistent with one another given the uncertainties.  We expect lower mass objects to have a smaller accretion disk size for a given effective wavelength, and the fact that the two are consistent could be due to the rather large mass uncertainties for all of these SMBHs, the small number of SMBHs at the lowest masses, or the unequal dynamic range probed by the mass bins due to our total sample size.

\section{Conclusion}
\label{sec: conc}
We report quasar accretion disk size measurements using time delays between the DES photometric bandpasses as a proxy for different radii in the disk.  In addition to modeling the individual pairs of photometric bands, we present a new JAVELIN tool that fits a thin disk model directly rather than to all the available light curves simultaneously, which we test on both NGC 5548 and our DES quasars.  Our results are:

\begin{enumerate}
    \item Even with the long cadence of the DES supernova pointings ($\sim$1 week) compared to the few light-day expected size of the accretion disk, we are able to place meaningful upper limits on lags between continuum bands using JAVELIN for many of our objects.  These limits are comparable to the sizes found for accretion disks through the gravitational lensing technique as well as other AGN and quasars that have disk sizes derived in photometric reverberation mapping studies.
    \item Our new extension of JAVELIN\footnote{Available along with the original JAVELIN and a quick tutorial at \url{https://bitbucket.org/nye17/javelin}.}, a thin disk model, is able to reproduce the AGN STORM result for NGC 5548 \citep{Fausnaugh16} by fitting for the thin disk parameters directly rather than each lag individually.  This new extension is publicly available as part of the JAVELIN software package and should have many applications to future, multi-wavelength time domain studies.
    \item When we fix the $\beta$ parameter at 4/3, we measure sizes for 15 DES quasars with this thin disk model.  The quasar sample spans almost two orders of magnitude in mass, and several of our measurements are of comparable precision to the disk lensing sizes.  Given our large uncertainties for most of our sample, the disk size measurements are consistent with both the larger disks of previous results \citep{Morgan10, Shappee15, Fausnaugh16, Jiang17} and with a thin disk accreting at a moderate rate compared to Eddington ($\sim$0.3).
    \item If quasars have a high intrinsic reddening, the larger disk sizes than those expected from thin disk predictions can be accounted for \citet{Gaskell17}.
    \item We have five quasars with variability in multiple DES observing seasons, and analyze each season of data independently with our thin disk model extension to JAVELIN.  In two of these quasars, we have variability in all three seasons.  The accretion disk sizes are consistent with each other in all of these cases.
\end{enumerate}

\section{Acknowledgements}
DM would like to thank Michael Fausnaugh for helpful discussions throughout the project and Martin Gaskell for providing useful feedback.
Funding for the DES Projects has been provided by the U.S. Department of Energy, the U.S. National Science Foundation, the Ministry of Science and Education of Spain,
the Science and Technology Facilities Council of the United Kingdom, the Higher Education Funding Council for England, the National Center for Supercomputing
Applications at the University of Illinois at Urbana-Champaign, the Kavli Institute of Cosmological Physics at the University of Chicago,
the Center for Cosmology and Astro-Particle Physics at the Ohio State University,
the Mitchell Institute for Fundamental Physics and Astronomy at Texas A\&M University, Financiadora de Estudos e Projetos,
Funda{\c c}{\~a}o Carlos Chagas Filho de Amparo {\`a} Pesquisa do Estado do Rio de Janeiro, Conselho Nacional de Desenvolvimento Cient{\'i}fico e Tecnol{\'o}gico and
the Minist{\'e}rio da Ci{\^e}ncia, Tecnologia e Inova{\c c}{\~a}o, the Deutsche Forschungsgemeinschaft and the Collaborating Institutions in the Dark Energy Survey.

The Collaborating Institutions are Argonne National Laboratory, the University of California at Santa Cruz, the University of Cambridge, Centro de Investigaciones Energ{\'e}ticas,
Medioambientales y Tecnol{\'o}gicas-Madrid, the University of Chicago, University College London, the DES-Brazil Consortium, the University of Edinburgh,
the Eidgen{\"o}ssische Technische Hochschule (ETH) Z{\"u}rich,
Fermi National Accelerator Laboratory, the University of Illinois at Urbana-Champaign, the Institut de Ci{\`e}ncies de l'Espai (IEEC/CSIC),
the Institut de F{\'i}sica d'Altes Energies, Lawrence Berkeley National Laboratory, the Ludwig-Maximilians Universit{\"a}t M{\"u}nchen and the associated Excellence Cluster Universe,
the University of Michigan, the National Optical Astronomy Observatory, the University of Nottingham, The Ohio State University, the University of Pennsylvania, the University of Portsmouth,
SLAC National Accelerator Laboratory, Stanford University, the University of Sussex, Texas A\&M University, and the OzDES Membership Consortium.

The DES data management system is supported by the National Science Foundation under Grant Number AST-1138766.
The DES participants from Spanish institutions are partially supported by MINECO under grants AYA2012-39559, ESP2013-48274, FPA2013-47986, and Centro de Excelencia Severo Ochoa SEV-2012-0234.
Research leading to these results has received funding from the European Research Council under the European Union’s Seventh Framework Programme (FP7/2007-2013) including ERC grant agreements
 240672, 291329, and 306478.

The data in this paper were based in part on observations obtained at the Australian Astronomical Observatory.

Part of this research was conducted by the Australian Research Council Centre of Excellence for All-sky Astrophysics (CAASTRO), through project number CE110001020.

Part of this research was funded by the Australian Research Council through project DP160100930.

This manuscript has been authored by Fermi Research Alliance, LLC under Contract No. DE-AC02-07CH11359 with the U.S. Department of Energy, Office of Science, Office of High Energy Physics. The United States Government retains and the publisher, by accepting the article for publication, acknowledges that the United States Government retains a non-exclusive, paid-up, irrevocable, world-wide license to publish or reproduce the published form of this manuscript, or allow others to do so, for United States Government purposes.

\bibliographystyle{mn2e}
\bibliography{continuum_lag_references}

\end{document}